\documentclass[letter]{article}

\usepackage[english]{babel}
\usepackage[utf8]{inputenc}
\usepackage[T1]{fontenc}

\usepackage[letterpaper,top=3cm,bottom=2cm,left=3cm,right=3cm,marginparwidth=1.75cm]{geometry}

\usepackage{amsmath}
\usepackage{graphicx}

\usepackage[colorlinks=true, allcolors=blue]{hyperref}
\usepackage{stackengine}
\usepackage{physics}
\usepackage{amssymb} 
\usepackage{mathrsfs}
\usepackage{amsfonts}
\usepackage{amssymb}
\usepackage[useregional]{datetime2}
\usepackage[left]{lineno}
\usepackage{longtable}
\usepackage{booktabs}
\usepackage[framemethod=TikZ]{mdframed}
\usepackage{tcolorbox}
\tcbuselibrary{breakable}
\usepackage{csquotes}
\usepackage{pdflscape}
\usepackage{tabularx}
\usepackage{longtable}
\usepackage{makecell}
\usepackage{adjustbox}
\usepackage{array}
\usepackage{csquotes}
\usepackage{float}
\usepackage{blindtext}
\usepackage[affil-it]{authblk}

\usepackage[backend=biber, maxcitenames=2, uniquename=false, style=apa, sorting=nyt,doi=false,isbn=false,url=false, natbib=true]{biblatex}

\DeclareNameFormat{labelname:poss}{
  \nameparts{#1}
  \ifcase\value{uniquename}%
    \usebibmacro{name:family}{\namepartfamily}{\namepartgiven}{\namepartprefix}{\namepartsuffix}%
  \or
    \ifuseprefix
      {\usebibmacro{name:first-last}{\namepartfamily}{\namepartgiveni}{\namepartprefix}{\namepartsuffixi}}
      {\usebibmacro{name:first-last}{\namepartfamily}{\namepartgiveni}{\namepartprefixi}{\namepartsuffixi}}%
  \or
    \usebibmacro{name:first-last}{\namepartfamily}{\namepartgiven}{\namepartprefix}{\namepartsuffix}%
  \fi
  \usebibmacro{name:andothers}%
  \ifnumequal{\value{listcount}}{\value{liststop}}{'s}{}}
\DeclareFieldFormat{shorthand:poss}{%
  \ifnameundef{labelname}{#1's}{#1}}
\DeclareFieldFormat{citetitle:poss}{\mkbibemph{#1}'s}
\DeclareFieldFormat{label:poss}{#1's}
\newrobustcmd*{\posscitealias}{%
  \AtNextCite{%
    \DeclareNameAlias{labelname}{labelname:poss}%
    \DeclareFieldAlias{shorthand}{shorthand:poss}%
    \DeclareFieldAlias{citetitle}{citetitle:poss}%
    \DeclareFieldAlias{label}{label:poss}}}
\newrobustcmd*{\posscite}{%
  \posscitealias%
  \textcite}
\newrobustcmd*{\Posscite}{\bibsentence\posscite}
\newrobustcmd*{\posscites}{%
  \posscitealias%
  \textcites}
 
\newcommand{\E}[2]{\mathbb{E}_{#1} \negmedspace \left[ #2 \right]}
\newcommand{\Cov}[3]{\mathrm{Cov}_{#1} \negmedspace \left( #2, #3 \right)}
\newcommand{\Var}[2]{\mathrm{Var}_{#1} \negmedspace \left( #2\right)}
\renewcommand{\eqref}[1]{Eq.\ref{#1}}
\newcommand{\tabitem}{~~\llap{\textbullet}~~}

\renewcommand\pdv[3]{\partialderivative{#1}{#2}{#3}{}}

\title{Methods for calculating coexistence mechanisms: Beyond scaling factors}
\date{\today}

\author[1,2,*]{Evan C. Johnson}
\author[1]{Alan Hastings}

\affil[1,]{Department of Environmental Science and Policy; University of California Davis; Davis, California 95616 USA}
\affil[2,]{Center for Population Biology; University of California Davis; Davis, California 95616 USA}
\affil[*]{Corresponding author: Evan Johnson, evcjohnson@ucdavis.edu}

\begin{document}
\maketitle
\clearpage

\section*{Abstract} 

How do species coexist? A framework known as Modern Coexistence Theory measures mechanisms of coexistence by comparing a species perturbed to low density (the \textit{invader}) to other species that remain at their typical densities (the \textit{residents}); this \textit{invader--resident comparison} measures a rare-species advantage that results from specialization. However, there are several reasonable ways (i.e., methods) to compare invaders and residents, each differing in practicality and biological interpretation. Here, using theoretical arguments and case studies, we compare four such methods for calculating coexistence mechanisms: 1) \textit{Scaling factors}, the traditional approach where resident growth rates are scaled by a measure of relative sensitivity to competition, obtained by solving a system of linear equations; 2) \textit{The simple comparison}, which gives equal weight to all resident species; 3) \textit{Speed conversion factors}, a novel method in which resident growth rates are scaled by a ratio of generation times, and; 4) \textit{The invader--invader comparison}, another novel method in which a focal species is compared to itself at high vs. low density. We conclude that the conventional scaling factors can be useful in some theoretical research, but are not recommended for empirical applications, i.e., determining the mechanisms of coexistence in real communities. Instead, we recommend the simple comparison and speed conversion factor methods. The speed conversion factors are most useful when comparing species with dissimilar generation times. However, ecologists often study coexistence in guilds of species with similar life-histories, and therefore, similar generation times. In such scenarios, the easier-to-use simple comparison method is reasonable. Our work demonstrates the importance of conceptual analysis: There is no self-evident way to define coexistence mechanisms, so one must justify a definition by showing that it corresponds to widely-held intuitions about what qualifies as a good explanation for coexistence. 

Keywords: Modern coexistence theory, scaling factors, comparison quotients, coexistence, invasion analysis, invader-invader comparison

\tableofcontents

\newpage

\section{Introduction} 
\label{Introduction}

Determining the mechanisms of coexistence in any given community is a difficult task. The underlying problem here is that nature is complex, but explanations for coexistence (being reductionistic, as all explanations are) are simple, often codified in simple two-species models and several paragraphs of commentary. Therefore, our task is to take a arbitrary, complex model (representing the real world), and extract the relative importance of several simple explanations. Modern Coexistence Theory is a tool that makes this possible.

Modern Coexistence Theory (MCT) is a framework for understanding coexistence. More specifically, MCT "measures coexistence" by quantifying \textit{coexistence mechanisms}: processes (e.g., resource specialization) that tend to increase species' per capita growth rates when rare. The ability of each species to recover from rarity is ostensibly related to the overall stability of the community -- coexistence. Crucially, coexistence mechanisms are operationalized with an \textit{invader--resident comparison}: processes that contribute the per capita growth rate of a species that has been perturbed to low density (the \textit{invader}) are compared to corresponding processes for species at typical abundances (the \textit{residents}).

However, to date, little attention has been paid to the interpretation of coexistence mechanisms. MCT is used unquestioningly in empirical applications, even though it was not originally designed to measure coexistence, but rather to produce theoretical insights about the role of fluctuations in coexistence (\cite[p.~288]{barabas2018chesson}; \cite[p.~6]{Chesson2019}). The absence of conceptual analysis is a problem: if our goal is to interpret the values of coexistence mechanisms as the relative importance of explanations for coexistence, then we must be sure of the correspondence between explanations and coexistence mechanisms. In other words, the exact definition of coexistence mechanisms is crucial to their interpretation, and thus crucial to how we understand coexistence.  

One part of the definition of coexistence mechanisms is \textit{scaling factors}, constants that re-scale the growth rates of residents. For many, the scaling factors are the most confusing part of MCT. They were introduced by \textcite[p.~241]{Chesson1994} with little justification: “This choice is justified by the results that it gives. It leads to a clear partitioning of mechanisms of coexistence, as shown in subsection 4.2, below." A diligent reader may go onto infer that the purpose of the scaling factors is to eliminate a term in the mathematical expression for the the invader's growth rate: "$\ldots$ linear terms in competition do not appear in this comparison $\ldots$" (\cite[p.~247]{Chesson1994}; also see equations 36 and 43). Decades later, \textcite[p.~3]{Chesson2019} confirms, "The idea [of the scaling factors] is that the [invader--resident] comparison should eliminate common components of competition to highlight critical species differences." One is left wondering why the linear effects of competition cannot be the basis of a critical species difference.

Further complicating the usage of scaling factor is that fact that they cannot serve their purported purpose --- eliminating the linear effects of competition --- when there are more distinct regulating factors than resident species (\cite{Chesson1994}; \cite{barabas2018chesson}). Regulating factors (also known as limiting factors or internal variables), are variables that are involved in a feedback loop that regulates population density. Examples include resources, refugia, and natural enemies. When a regulating factor is \textit{continuous} (e.g., seeds along a continuum of sizes, varying rates of resource supply across space), then there are technically an infinite number of regulating factors, and thus the scaling factors automatically cannot serve their purported purpose.

There are a slew of other problems with scaling factors. 1) When species' sensitivities to regulating factors are similar, small differences in species'  sensitivities will lead to big differences in the scaling factors (due to inverting an ill-conditioned matrix; see \cite[p.~295]{barabas2018chesson}). This means that inferences from empirical applications of MCT can be sensitive to measurement error and/or parameter estimation error. 2) Scaling factors can switch from positive to negative, turning an invader--resident difference into a invader--resident sum (due to subtracting a negative; \cite[p.~E92]{snyder2005examining}); this is problematic because the invader--resident difference is what permits us to interpret coexistence mechanisms as a rare-species advantage. 3) When a certain assumption of the mathematical theory is not met (Assumption \textit{a6} in \cite{Chesson1994}), the scaling factors may not be uniquely determined, even when there are more residents than regulating factors. In this scenario, the analytic theory cannot be used to calculate scaling factors, and instead one must use one of several computationally-intensive work-arounds (see \cite{ellner2016quantify}, SI.5). 

Here, we argue that the primary function of the scaling factors --- eliminating the linear effects of competition --- is not desirable if one wants to use MCT to understand coexistence in real communities. Eliminating the linear effects of competition is effective at showing that not all species can coexist via classical mechanisms (i.e., fluctuation-independent mechanisms such as resource or natural-enemy partitioning), which can be useful in theoretical research. But, if one wants to "measure coexistence" (i.e., understand empirically how species coexist) then is desirable to be able to attribute coexistence to classical mechanisms.   

However, it is not merely the case that scaling factors are unnecessary: they can also lead to incorrect inferences about how species are coexisting. Scaling factors are designed entirely to eliminate the linear effects of competition, but they have an auxiliary usage: to weight resident growth rates in the calculation of other coexistence mechanisms. Therefore (as we will show), the scaling factors can modulate other coexistence mechanisms, sometimes in a way that is nonsensical.

The obvious alternative to scaling factors is a simple average over resident species, which which we call the \textit{simple comparison} method. To be more precise, each resident gets weighted by $1/(S-1)$ (where $S$ is the number of species in the community), such that equal weight is given to the low density state (i.e., the invader) and the high density state (i.e., the sum of residents). The simple comparison can work well, but it can also be problematic if some species have comparatively fast population dynamics; such species tend to dominate all other species in the invader--resident comparison.

Our proposed solution to the shortcomings of previous methods is to scale resident growth rates by \textit{speed conversion factors}. Conceptually, speed conversion factors convert the intrinsic speed of resident population dynamics to that of the invader. Functionally, the speed conversion factors prevent the invader--resident comparison from being dominated by terms corresponding to a handful of speedy species. Yet another solution is to replace the invader--resident comparison with an \textit{invader--invader comparison}, wherein a single focal species is compared to itself at high vs. low density. 

In this paper, we define and discuss the four aforementioned methods for calculating coexistence mechanisms: scaling factors (Section \ref{Scaling factors}), the simple comparison (Section \ref{The simple comparison}), speed conversion factors (Section \ref{Speed conversion factors}), and the invader--invader comparison (Section \ref{The invader--invader comparison}). We discuss the strengths and weakness of each method (see Table \ref{procon} in the \textit{Discussion}), using both conceptual arguments and case studies (Section \ref{Case studies}).

\section{What are scaling factors?}
\label{What are scaling factors?}

\subsection{A summary of Modern Coexistence Theory}
\label{A summary of Modern Coexistence Theory}

To understand scaling factors, one must first understand the basics of Modern Coexistence Theory (MCT). In MCT, a rare-species advantage is operationalized as an \textit{invasion growth rate}: the long-term average of the per capita growth rate of the invader. The main innovation of MCT is the partition of the invasion growth rate into \textit{coexistence mechanisms}. This partition is obtained with two main steps: "decompose and compare" (\cite{Ellner2019}).

\begin{enumerate}
    \item \textbf{Decompose}.

    Consider a community composed of scalar populations (i.e., populations without age, stage, or spatial structure), subject to temporal variation in the environment, population densities, and regulating factors. The per capita growth rate of species $j$ is denoted by $r_j(t) = dN_j(t)/(N_j(t) dt)$ in continuous time or $r_j(t) = \log(N_j(t+1) / N_j(t))$ in discrete time. Now, we perturb species $i$ (the invader) to zero density and use the superscript "$\{-i\}$" to indicate quantities that must be evaluated in this context. We represent the per capita growth rate as a function $g_j$ of the environmental parameter $E_j(t)$, as well as $L$ regulating factors, $\boldsymbol{F^{\{-i\}}(t)} = (F_1^{\{-i\}}(t), F_2^{\{-i\}}(t), ..., F_L^{\{-i\}}(t))$:
    
    \begin{equation}
        r_j^{\{-i\}}(t) = g_j(E_j(t), \boldsymbol{F^{\{-i\}}(t)}).
    \end{equation}
    
    For notational simplicity, we will drop the explicit time-dependence. The parameter $E_j$ is sometimes called the environmental response, or the environmentally-dependent parameter, or simply the environment. While it usually represents a demographic parameter that depends on the environment (e.g., the probability of seed germination), it more generally represents the influence of density-independent factors. The regulating factors $\boldsymbol{F}$ can be abiotic resources, biotic resources, species densities, natural enemies, refugia, light, etc.

    Next, we approximate each species' per capita growth rate with a second order Taylor series expansion about the equilibrium values $E_{j}^*$ and $\boldsymbol{F}^{*j}$, selected so that $r_j(E_{j}^*, \boldsymbol{F}^{*j}) = 0$. While the regulating factors $\boldsymbol{F}$ are not species-specific, the equilibrium values  $\boldsymbol{F}^{*j} = (F_1^{*j}, F_2^{*j}, ..., F_L^{*j})$ may be species-specific (not the superscript "$j$"). There is no agreed upon method for determining the equilibrium parameters, but they should be close to their respective temporal means, $\overline{E_{j}}$ and $\overline{\boldsymbol{F^{\{-i\}}}}$, in order for the Taylor series to be a good approximation (\cite[p.~280]{barabas2018chesson}). 
    
    In the case of one regulating factor, the canonical way to select the equilibrium parameters is to set environmental noise to zero (thus creating a \textit{deterministic skeleton}), set $E_j^*$ as the now-fixed environmental parameter, and then solve for $F^{*j}$ (see \cite[Section~5]{Chesson1994}). Alternatively, \textcite{barabas2018chesson} suggests selecting $E_j^* = \overline{E_j}$ (without first eliminating environmental noise), and then solving for $F^{*j}$. In the case of multiple regulating factors, \textcite{Chesson2019} suggests selecting reasonable $\boldsymbol{F}^{*j}$ first, and then solving for $E_j^*$. For instance, one could simulate the full community dynamics (where all species are present), set the equilibrium regulating factors to their temporal averages, $\overline{\boldsymbol{F}}$, and then solve for $E_j^*$ for each species; but because the quality of the Taylor series approximation of the invasion growth rate depends on $\boldsymbol{F}^{*j}$ being close to $\overline{\boldsymbol{F}^{\{-i\}}}$ (the temporal averages of regulating factors in the community without invader $i$), this method could work poorly if putting a species in the invader state would substantially change the mean levels of the regulating factors. We conclude that the equilibrium parameters must be chosen on a case-by-base basis. 
    
    After calculating the aforementioned Taylor series expansion, we take a temporal average in order to obtain an approximation of species $j$'s \textit{long-term} average growth rate. The result is

\begin{equation} \label{taylor_decomp_avg1}
\begin{aligned}
\overline{r_{j}^{\{-i\}}} \approx \; & \alpha_j^{(1)} (\overline{E_j} - E_{j}^{*}) + \frac{1}{2} \alpha_j^{(2)} \Var{}{E_j} \\ & + \sum_{k = 1}^{L} \phi_{jk}^{(1)} (\overline{F_k^{\{-i\}}} - F_{k}^{*j}) \\ & + \sum_{k = 1}^{L} \sum_{m = 1}^{L} \phi_{jkm}^{(2)} \Cov{}{F_k^{\{-i\}}}{F_m^{\{-i\}}} \\ &
 + \zeta_{jk}^{(1)}  \sum_{k = 1}^{L} \Cov{}{E_j}{F_k^{\{-i\}}},
\end{aligned}
\end{equation}

where the coefficients of the Taylor series, 

\begin{equation}  \label{taylor_coef}
\begin{aligned}
 \alpha_j^{(1)} = \pdv{g_j\scriptstyle{(E_j^*, \boldsymbol{F}^{*j})}}{E_j},  \quad
 \phi_{jk}^{(1)} = \pdv{g_j\scriptstyle{(E_j^*, \boldsymbol{F}^{*j})}}{F_k},  \quad
 \alpha_{j}^{(2)} = \pdv{g_j\scriptstyle{(E_j^*,\boldsymbol{F}^{*j})}}{E_j},  \quad
 \phi_{jkm}^{(2)} = \pdv{g\scriptstyle{(E_j^*, \boldsymbol{F}^{*j})}}{F_k}{F_m},  \quad
 \zeta_{jk}^{(1)} = \pdv{g\scriptstyle{(E_j^*, \boldsymbol{F}^{*j})}}{E_j}{F_k},  \quad
\end{aligned}
\end{equation}

are all evaluated at $E_j = E_j^*$ and $\boldsymbol{F} = \boldsymbol{F}^{*j}$. The quality of this approximation depends on the environmental parameter only experiencing small deviations from equilibrium (for the mathematical details, see \textcite{Chesson1994}; \textcite{chesson2000general}). These \textit{small-noise assumptions} also allow us to replace $\overline{(E_j-E_j^*)(F_k^{\{-i\}}-F_k^{*j})}$ with $\Cov{}{E_j}{F_k^{\{-i\}}}$ and perform analogous replacements for other terms.

    \item \textbf{Compare}
    
Resident species (denoted with subscript $s$) have a long-term average per capita growth rate of zero; otherwise, resident populations would go extinct or explode to infinity. Therefore, the value of the invasion growth rate is unaltered if we subtract a linear combination of the residents' average growth rates. That is, we can write the invasion growth rate of species $i$ as

\begin{equation} \label{inv res comparison}
    \overline{r_i^{\{-i\}}} = \overline{r_i^{\{-i\}}} - \sum \limits_{s \neq i}^S q_{is}  \overline{r_s^{\{-i\}}},
\end{equation}

where the $q_{is}$ are the scaling factors, and $S$ is the number of total species in the community. To identify the processes that generate a rare-species advantage, we can substitute \eqref{taylor_decomp_avg1} into \eqref{inv res comparison} and group like-terms. The invasion growth rate now becomes

\begin{equation} \label{MCT scaling factors}
\begin{aligned}
    \overline{r_i^{\{-i\}}} \approx & \underbrace{\alpha_i^{(1)} (\overline{E_i} - E_{i}^{*}) + \frac{1}{2} \alpha_i^{(2)} \Var{}{E_i} -  \left( \sum_{k = 1}^{L} \phi_{ik}^{(1)} F_{k}^{*i} \right)  - \sum \limits_{s \neq i}^S q_{is} \left((\overline{E_s} - E_{s}^{*}) + \frac{1}{2} \alpha_s^{(2)} \Var{}{E_s} - \sum_{k = 1}^{L} \phi_{sk}^{(1)} F_{k}^{*s} \right) }_{r'_{i} \colon \text{Density-independent effects}}  \\ 
    + & \underbrace{ \left( \sum_{k = 1}^{L} \phi_{ik}^{(1)} \overline{F_k^{\{-i\}}} \right) - \sum \limits_{s \neq i}^S q_{is}  \left( \sum_{k = 1}^{L} \phi_{sk}^{(1)} \overline{F_k^{\{-i\}}} \right) }_{\Delta \rho_i \colon \text{Linear density-dependent effects}} \\
    + & \underbrace{\frac{1}{2}\left[ \left( \sum_{k = 1}^{L} \sum_{m = 1}^{L} \phi_{ikm}^{(2)} \Cov{}{F_k^{\{-i\}}}{F_m^{\{-i\}}} \right)  - \sum \limits_{s \neq i}^S q_{is}   \sum_{k = 1}^{L} \sum_{m = 1}^{L} \phi_{skm}^{(2)} \Cov{}{F_k^{\{-i\}}}{F_m^{\{-i\}}} \right]}_{\Delta N_i \colon \text{Relative nonlinearity}} \\
    + & \underbrace{\sum_{k = 1}^{L} \zeta_{ik}^{(1)} \Cov{}{E_i}{F_k^{\{-i\}}}   - \sum \limits_{s \neq i}^S q_{is}  \sum_{k = 1}^{L} \zeta_{sk}^{(1)} \Cov{}{E_s}{F_k^{\{-i\}}}}_{\Delta I_i\ \colon \text{The storage effect}}
\end{aligned}
\end{equation}

The symbols under the brackets ($r'_i$, $\Delta \rho_i$, $\Delta N_i$, and $\Delta I$) denote the coexistence mechanisms.

The interpretations of the coexistence mechanisms are as follows: \textit{The density independent effects}, $\Delta E_i$, is the degree to which all density-independent factors favor the invader. \textit{The linear density-dependent effects}, $\Delta \rho_i$, represents a rare-species advantage due to specialization on regulating factors (e.g., resources, natural enemies). \textit{Relative nonlinearity}, $\Delta N_i$, is a rare-species advantage due to specialization on variation in regulating factors. \textit{The storage effect}, $\Delta I_i$, is the rare-species advantage due to specialization on certain states of a variable environment. See \textcite{barabas2018chesson} for a more thorough discussion of the coexistence mechanisms, their interpretations, and their connection to specific models.

\end{enumerate}

Experts may note that our presentation of MCT differs subtly from that of previous research. First, our exposition above only accommodates models with temporal variation. There is an analogous version of MCT for models with spatial variation (\cite{chesson2000general}), which we do not present here for the sake of simplicity. Second, we write the per capita growth rate directly as a function of shared (across species) regulating factors, as opposed to a function of a species-specific competition parameter, which itself a function of the shared regulating factors (as in \cite{Chesson1994}; \cite{barabas2018chesson}). Third, we do not change coordinates to the so-called \textit{standard parameters}, $\mathscr{E}_j$ and $\mathscr{C}_j$ (see \cite{Chesson1994}, p. 236). In our opinion, the standard parameters impose another layer of abstraction that makes MCT more difficult to understand, particularly because previous work includes formulas that mix the standard parameters with the normal parameters, $E_j$ and $C_j$ (e.g., Eq.19 in \cite{barabas2018chesson}).

The main function of the standard parameters, $\mathscr{E}_j$ and $\mathscr{C}_j$, is to generate coexistence mechanisms that sum exactly to the invasion growth rate (\cite{Chesson2019}), which we call \textit{exact coexistence mechanisms}. The exact coexistence mechanisms are calculated in our case studies and corresponding Mathematica notebooks, but because they are less conventional and more difficult to define succinctly, here we only present the approximate coexistence mechanisms.

\subsection{Scaling factors}
\label{Scaling factors}

The scaling factors were introduced by \textcite{Chesson1994}, but were referred to only by the symbols $q_{is}$; only recently did \textcite{ellner2016quantify} coin the term \textit{scaling factors}. Chesson (\citeyear{Chesson2018}, \citeyear{Chesson2019}) suggests the term \textit{comparison quotients}, but we use scaling factors here purely for the sake of convention. 

First, we define the standard competition parameter:
\begin{equation} \label{def}
        \mathscr{C}_j(t) = g_j(E_j^*, \boldsymbol{F}). 
\end{equation}    
The regulating factors may have different meanings in different models (representing different species and/or communities); it may even have different units (e.g., abundance vs. biomass). By contrast, the standard competition parameter is always defined with the the common currency of growth rates (or pseudo-rates in the case of discrete time), and can therefore be thought of as the main effect of competition on the average per capita growth rate.

The scaling factors are defined as 
\begin{equation} \label{q_ir_chessons_def}
    q_{is} = \frac{\partial \mathscr{C}_{i}^{\{-i\}}}{\partial \mathscr{C}_{r}^{\{-i\}}},
\end{equation}
evaluated at $\mathscr{C}_{r} = 0$, or equivalently, $\boldsymbol{F} = \boldsymbol{F}^{*r}$. 

Next, we make two assumptions (respectively Assumption \textit{a6} and Eq.49 in \cite{Chesson1994}). Assume that we can express the standard competition parameter of the invader as a function of the residents' standard competition parameters:
\begin{equation} \label{a6}
    \mathscr{C}_{i}^{\{-i\}} = f \left(\mathscr{C}_{1}^{\{-i\}}, \ldots, \mathscr{C}_{S}^{\{-i\}} \right).
\end{equation}
Further, assume that the standard competitive parameters can be written as a function of $L$ competitive factors $(F_1, \ldots, F_L)^\intercal = \boldsymbol{F}$:

\begin{equation}
     \mathscr{C}_{j} = \phi_j(F_1, \ldots, F_L) = \phi_j(\boldsymbol{F}).
\end{equation}

The competitive factors can now be related to the scaling factors through the chain rule,
\begin{equation} \label{limiting_factors_to_qir}
    \frac{\partial \mathscr{C}_{i}^{\{-i\}} }{F_k} = \sum \limits_{r \neq i}^S \frac{\partial \mathscr{C}_{i}^{\{-i\}}}{\partial \mathscr{C}_{r}^{\{-i\}}} \frac{\partial \mathscr{C}_{r}^{\{-i\}} }{F_k}
\end{equation}
with all derivatives evaluated at $\mathscr{C}_{r}^{\{-i\}} = 0$ and $\boldsymbol{F} = \boldsymbol{F^{*r}}$. The partial derivatives of $\mathscr{C}_{j}^{\{-i\}}$ with respect to $F_k$ are the first order coefficients of a Taylor series of $\phi_j$ expanded about  $\boldsymbol{F} = \boldsymbol{F^{*j}}$, so we we may use the notation  $\phi_{jk}^{(1)} = \left. \partial \mathscr{C}_{j} / {\partial F_k} \right|_{F_k = F_{k}^{*j}} = \left. \partial \mathscr{C}_{j}^{\{-i\}} / {\partial F_k} \right|_{F_k = F_{k}^{*j}}$
  
Substituting $\phi_{jk}^{(1)}$ and the left-hand-side of \eqref{def} into \eqref{limiting_factors_to_qir}, we get
\begin{equation} \label{limiting_factors_to_qir_new_notation}
    \phi_{ik}^{(1)} = \sum \limits_{r \neq i}^S q_{ir} \phi_{rk}^{(1)}.
\end{equation}
With this one equation and $S-1$ unknowns (the $q_{ir}$'s), \eqref{limiting_factors_to_qir_new_notation} is underdetermined. If we consider the equations of all $L$ competitive factors simultaneously, we get the vector-matrix equation
\begin{equation} \label{vector_matrix_qir}
    \boldsymbol{\Phi}_{i*} = \boldsymbol{q_{i*}} \boldsymbol{\Phi}^{(-i)}.
\end{equation}
Here, $\boldsymbol{\Phi}$ is a ($S \times L$) matrix of species' sensitivities to competitive factors, with elements $\phi_{jk}^{(1)}$. The symbol $\boldsymbol{\Phi}_{i*}$ is the ($1 \times L$) row vector of the invader's sensitivities to regulating factors; $\boldsymbol{q_{i*}}$ is the ($1 \times (S-1)$) row vector of scaling factors (the element $q_{ii}$ is not included); and $\boldsymbol{\Phi}^{(-i)}$ is a ($(S-1) \times L$) matrix, obtained by removing the invader (i.e., the $i$-th row) from $\boldsymbol{\Phi}$. 

Solving for $\boldsymbol{q_{i*}}$ involves multiplying both sides of \eqref{vector_matrix_qir} by the inverse of $\boldsymbol{\Phi}^{(-i)}$. However, the invertible matrix theorem states that the matrix inverse only exists if $\boldsymbol{\Phi}^{(-i)}$ is square (i.e., $S-1 = L$) and has linearly independent columns. What do we do when $\boldsymbol{\Phi}^{(-i)}$  is not square: when there are more residents than regulating factors, or more regulating factors than residents?

\posscite{Chesson1994} solution is the \textit{generalized inverse}. A generalized inverse of a matrix  $\boldsymbol{A}$ is denoted as $\boldsymbol{A}^g$, and satisfies the equation $\boldsymbol{A} \boldsymbol{A}^g \boldsymbol{A} = \boldsymbol{A}$ 
(\cite{ben2003generalized}). Our expression for the scaling factors now becomes

\begin{equation} \label{qir_sol}
    \boldsymbol{q_{i*}} = \boldsymbol{\Phi}_{i*} \left( \boldsymbol{\Phi}^{(-i)} \right)^g.
\end{equation}

When $\boldsymbol{\Phi}^{(-i)}$ is square and has full rank, then the regular matrix inverse is the unique generalized inverse. When the linear system is underdetermined (i.e., $S-1 > L$, assuming $\boldsymbol{\Phi}^{(-i)}$ has full rank), then a generalized inverse produces an infinite number of solutions to \eqref{vector_matrix_qir}. According to Chesson (\citeyear{Chesson2019}, p. 6) this non-uniqueness is a virtue: different choices of generalized inverses allow the user of MCT to ask different scientific questions. 

When a solution is available, the resulting scaling factors (\eqref{qir_sol}) can be used to cancel the linear effects of density-dependence (i.e., $\Delta \rho_i$ in \eqref{MCT scaling factors}). The  linear effects of density-dependence can be expressed in vector-matrix form:
\begin{equation} 
    \Delta \rho_i = \boldsymbol{\Phi}_{i*} \overline{\boldsymbol{F^{\{-i\}}}} - \boldsymbol{q_{i*}} \boldsymbol{\Phi}^{(-i)} \overline{\boldsymbol{F^{\{-i\}}}}.
\end{equation}
Substituting in the right-hand-side of \eqref{qir_sol}, we get
\begin{equation} \label{qir_almost_cancel}
    \Delta \rho_i = \boldsymbol{\Phi}_{i*} \overline{\boldsymbol{F^{\{-i\}}}} - \boldsymbol{\Phi}_{i*} \left( \boldsymbol{\Phi}^{(-i)} \right)^g \boldsymbol{\Phi}^{(-i)} \overline{\boldsymbol{F^{\{-i\}}}}.
\end{equation}
The matrix product $\left( \boldsymbol{\Phi}^{(-i)} \right)^g \boldsymbol{\Phi}^{(-i)}$ evaluates to the ($L \times L$) identity matrix, and therefore, $\Delta \rho_i = 0$.

When the linear system in \eqref{vector_matrix_qir} is overdetermined (i.e., $S-1 < L$, assuming $\boldsymbol{\Phi}^{(-i)}$ has full rank), \eqref{qir_sol} can still be used, but the resulting $\boldsymbol{q_{i*}}$ will not be a strict solution. \textcite{barabas2018chesson} suggests cancelling a major regulating factor that has a particularly strong effect on per capita growth rates. In fact, one could cancel up to $(L-S+1)$ such major regulating factors by computing the generalized inverse for a submatrix of $\boldsymbol{\Phi}^{(-i)}$ that contains only columns corresponding to the major regulating factors. 

In this paper, we will argue against most uses of the scaling factors. However, if one still desires to use the scaling factors, we suggest using the Moore-Pensrose Pseudoinverse (denoted with a dagger: $\dagger$) when the linear system in \eqref{vector_matrix_qir} is underdetermined or overdetermined. Specifically, in the case of overdetermination, we suggest

\begin{equation} \label{qir_approx_sol_over}
    \boldsymbol{q_{i*}} = \boldsymbol{\Phi}_{i*} \left( \boldsymbol{\Phi}^{(-i)} \right)^\dagger.
\end{equation}

The pseudoinverse gives the optimal solution in the least-squares sense (\cite[p.~122]{ben2003generalized}), so while it may be impossible to cancel $\Delta \rho_i$, it may be possible to get close. In the case of underdetermination, we suggest

\begin{equation} \label{qir_approx_sol_under}
    \boldsymbol{q_{i*}} = \boldsymbol{z} + (\boldsymbol{\Phi}_{i*} - \boldsymbol{z} \boldsymbol{\Phi}^{(-i)} ) \left( \boldsymbol{\Phi}^{(-i)} \right)^\dagger,
\end{equation}

where $\boldsymbol{z}$ is an $(1 \times (S-1) )$ row vector where each element is equal to $1/(S-1)$. This formula comes from taking \eqref{vector_matrix_qir}, replacing $\boldsymbol{q_{i*}}$ with $\boldsymbol{x} + \boldsymbol{z}$, and attempting to solve for $\boldsymbol{x}$. The answer gives the minimum norm solution (\cite[p.~109]{ben2003generalized}) for $\boldsymbol{x}$, which means $\boldsymbol{q_{i*}} = \boldsymbol{x} + \boldsymbol{z}$ is close $\boldsymbol{z}$ in the least-squares sense. In other words, we eliminate $\Delta \rho_i$ while deviating a minimal amount from the simple average over residents.

\subsection{When are scaling factors useful? When are they not useful?}
\label{When are scaling factors useful? When are they not useful?}

The purpose of the scaling factor is to eliminate $\Delta \rho_i$. But why eliminate $\Delta \rho_i$? To our knowledge, the most explicit explanation comes from \textcite[3]{Chesson2019}: "The idea is that the [invader--resident] comparison should eliminate common components of competition to highlight critical species differences." Again, it is not clear why $\Delta \rho_i$ is not a critical species difference, particularly since it encapsulates classical explanations for species coexistence: resource partitioning and natural-enemy partitioning. Perhaps we can arrive at a clearer justification of the scaling factors by studying papers in which the scaling factors played a crucial role.

\textcite{chesson1997roles} analyzed a model where per capita growth rates responded linearly to environmental fluctuations and a single regulating factor. The scaling factors eliminated $\Delta \rho_i$, and the linear responses precluded the fluctuation-dependent mechanisms, $\Delta N_i$ and $\Delta I_i$. Thus, a species' average growth rate could be represented entirely by the density-independent effects, $r_i'$. One can show that a weighted sum of $r_i'$ across species is equal to zero; if some species have a positive $r_i'$, others necessarily have a negative $r_i'$, so at least one species is destined for extinction. This result is a triumph of the scaling factors because it contradicted the idea that disturbances per se promoted coexistence (\cite{wiens1977competition}; \cite{huston1979general}; \cite{strong1983natural}).

The scaling factors can also highlight the role of fluctuations in coexistence. If $\Delta \rho_i$ is cancelled, then not all species can coexist on $r_i'$. If species nonetheless coexist, then coexistence must be attributable to fluctuation-dependent mechanisms, $\Delta N_i$ and/or $\Delta I_i$. Using this approach, \textcite{Chesson1994} showed that fluctuations are necessary for coexistence in the the lottery model and the annual plant model. Crucially, in both of the aforementioned papers (\cite{chesson1997roles}; and \cite{Chesson1994}), the cancelling of $\Delta \rho_i$ is valuable because it tells us how species are \textit{not} coexisting. 

There is also one completely pragmatic reason for canceling $\Delta \rho_i$: with this term cancelled, we do not need to calculate the temporal averages of the regulating factors, $\overline{\boldsymbol{F^{\{-i\}}}}$. This is mainly useful for theorists, since analytical expressions for $\overline{\boldsymbol{F^{\{-i\}}}}$ can be unobtainable, or too complicated to be insightful. By contrast, it is always possible to numerically evaluate $\overline{\boldsymbol{F^{\{-i\}}}}$. 

Given that the scaling factors are a seminal part of MCT, which itself is an all-purpose framework, it is easy to get the impression that the scaling factors are also all-purpose. Yet the historical record shows that the scaling factors have been used as a means specific ends: expanding on the competitive exclusion principle, highlighting the role of fluctuation-dependent mechanisms, and simplifying mathematical formulas. More generally, the scaling factors are suited for deriving biological insights from the mathematical analysis of simple models. The original goal of MCT was to understand how fluctuations affect coexistence (\cite[p.~288]{barabas2018chesson}, \cite[p.~6]{Chesson2019}), and indeed, the scaling factors have proved valuable in pursuit of this goal. 

However, there is growing interest in using MCT as a measurement tool; as a way to quantify the mechanisms of coexistence in real communities (e.g., \cite{caceres1997temporal}; \cite{adler2006climate}; \cite{sears2007new}; \cite{descamps2005stable}; \cite{angert2009functional}; \cite{Adler2010}; \cite{usinowicz2012coexistence}; \cite{chesson2012storage}; \cite{chu2015large}; \cite{usinowicz2017temporal}; \cite{hallett2019rainfall}; \cite{armitage2019negative}; \cite{armitage2020coexistence}; \cite{zepeda2019fluctuation}; \cite{zepeda2019fluctuation}; \cite{towers2020requirements}; \cite{ellner2016quantify}; \cite{Ellner2019}). It is arguable, \textit{a priori}, that the scaling factors do not serve this goal. We want to know the degree to which classical explanations  (i.e., resource and natural-enemy partitioning) promote coexistence, so we should not try to cancel $\Delta \rho_i$. Quite the opposite --- to gain a more fine-grained understanding of coexistence, we typically expand $\Delta \rho_i$ into contributions from individual regulating factors.

It is not merely the case that scaling factors are unnecessary; in certain cases, they can impede our understanding of coexistence. In a case study (Section \ref{Case studies}), we work through an example where it is intuitively clear that species 1 is specializing on resource 1, species 2 is specializing on resource 2, and species 3 is specializing on the variation in resource 2. Following the biological interpretations of the coexistence mechanisms in Section \ref{A summary of Modern Coexistence Theory} (see \cite{barabas2018chesson} for more thorough interpretations), it is intuitively clear that $\Delta \rho_1$ should be large and positive, $\Delta \rho_2$ should be large and positive, and $\Delta N_3$ should be large and positive. However, because of the scaling factors, MCT tells us that species 1 is persisting via relative nonlinearity ($\Delta N_1$ is large and positive), despite the fact that species 1 has a linear response to resource levels and barely interacts with the other species. The problem is that scaling factors are singularly focused on canceling $\Delta \rho_i$, which may cause unwanted collateral modulations to other coexistence mechanisms. 

\section{Alternatives to scaling factors}
\label{Alternatives to scaling factors}

\subsection{The simple comparison}
\label{The simple comparison}

\textcite{Ellner2019} suggested abandoning the scaling factors and redefining the coexistence mechanisms with a simple average over resident species. This leads to the following partition of the invasion growth rate:
\begin{equation} \label{MCT simple}
\begin{aligned}
    \overline{r}_i \approx & \underbrace{ \left(\alpha_i^{(1)} (\overline{E_i} - E_{i}^{*}) + \frac{1}{2} \alpha_i^{(2)} \Var{}{E_i}\right)  - \frac{1}{S-1} \sum \limits_{s \neq i}^S  \left((\overline{E_s} - E_{s}^{*}) + \frac{1}{2} \alpha_s^{(2)} \Var{}{E_s} \right) }_{\Delta E_{i} \colon \text{Density-independent effects}}  \\ 
    + & \underbrace{ \left( \sum_{k = 1}^{L} \phi_{ik}^{(1)} \left(\overline{F_k^{\{-i\}}} - F_k^{*i} \right) \right) - \frac{1}{S-1} \sum \limits_{s \neq i}^S  \left( \sum_{k = 1}^{L} \phi_{sk}^{(1)} \left(\overline{F_k^{\{-i\}}} - F_k^{*s} \right) \right) }_{\Delta \rho_i \colon \text{Linear density-dependent effects}} \\
    + & \underbrace{\frac{1}{2}\left[ \left( \sum_{k = 1}^{L} \sum_{m = 1}^{L} \phi_{ikm}^{(2)} \Cov{}{F_k^{\{-i\}}}{F_m^{\{-i\}}} \right)  - \frac{1}{S-1} \sum \limits_{s \neq i}^S  \left(  \sum_{k = 1}^{L} \sum_{m = 1}^{L} \phi_{skm}^{(2)} \Cov{}{F_k^{\{-i\}}}{F_m^{\{-i\}}} \right) \right]}_{\Delta N_i \colon \text{Relative nonlinearity}} \\
    + & \underbrace{\left(\sum_{k = 1}^{L} \zeta_{ik}^{(1)} \Cov{}{E_i}{F_k^{\{-i\}}} \right)   - \frac{1}{S-1} \sum \limits_{s \neq i}^S \left( \sum_{k = 1}^{L} \zeta_{sk}^{(1)} \Cov{}{E_s}{F_k^{\{-i\}}} \right). }_{\Delta I_i\ \colon \text{The storage effect}}
\end{aligned}
\end{equation}
Note that because $\Delta \rho_i$ does not need to be cancelled, there is no need to shunt $\boldsymbol{F}^{*j}$ terms from $\Delta \rho_i$ to the density-independent effects. Because the density-independent effects contains only environmental parameters, it is now denoted by $\Delta E_i$.

We can further expand $\Delta \rho_i$ into contributions from individual regulating factors. For instance, the degree to which species $i$ specializes on regulating factor $k$ is
\begin{equation}
    \Delta \rho_{i,F_{k}} = \left(\phi_{ik}^{(1)} \left(\overline{F_k^{\{-i\}}} - F_k^{*i} \right) \right) - \frac{1}{S-1} \sum \limits_{s \neq i}^S  \left( \phi_{sk}^{(1)} \left(\overline{F_k^{\{-i\}}} - F_k^{*s} \right) \right).
\end{equation}
If one so desires, similar expansions could be applied to the other consistence mechanisms. 

Coexistence is understood as a rare-species advantage, which necessitates a comparison of low-density states and high-density states. The simple comparison ostensibly gives equal weight to the low-density state and the high density state, while equally utilizing each resident. However, there is a sense in which the simple factors do not give equal weight to all residents. Consider a single resident that has the capacity to grow and decline at a rapid rate. Even though the resident's average growth rate is zero, the resident's grow rate components (i.e., the additive terms in \eqref{taylor_decomp_avg1}) will tend to be large in magnitude, and will therefore tend to dominate the invader--resident comparison. The simple comparison inappropriately emphasizes species with fast life-cycles.

\subsection{Speed conversion factors}
\label{Speed conversion factors}

When some species have the capacity to grow much more quickly that others, we recommend scaling resident growth rates by \textit{speed conversion factors}. The speed conversion factors are denoted $a_i / a_s$, where $a_j$ is a constant that represents the intrinsic speed of species $j$'s population dynamics. The terminology \textit{speed conversion factor} comes from the fact that $a_s$ in the denominator is cancelled by the $a_s$ implicit in the a resident's growth rate, leaving only the invader's speed, $a_i$.

Using speed conversion factors, the partition of the invasion growth rate becomes

\begin{equation} \label{MCT speed}
\begin{aligned}
    \overline{r}_i \approx & \underbrace{ \left(\alpha_i^{(1)} (\overline{E_i} - E_{i}^{*}) + \frac{1}{2} \alpha_i^{(2)} \Var{}{E_i}\right)  - \frac{1}{S-1} \sum \limits_{s \neq i}^S  \frac{a_i}{a_s} \left((\overline{E_s} - E_{s}^{*}) + \frac{1}{2} \alpha_s^{(2)} \Var{}{E_s} \right) }_{\Delta E_{i} \colon \text{Density-independent effects}}  \\ 
    + & \underbrace{ \left( \sum_{k = 1}^{L} \phi_{ik}^{(1)} \left(\overline{F_k^{\{-i\}}} - F_k^{*i} \right) \right) - \frac{1}{S-1} \sum \limits_{s \neq i}^S \frac{a_i}{a_s} \left( \sum_{k = 1}^{L} \phi_{sk}^{(1)} \left(\overline{F_k^{\{-i\}}} - F_k^{*s} \right) \right) }_{\Delta \rho_i \colon \text{Linear density-dependent effects}} \\
    + & \underbrace{\frac{1}{2}\left[ \left( \sum_{k = 1}^{L} \sum_{m = 1}^{L} \phi_{ikm}^{(2)} \Cov{}{F_k}{F_m} \right)  - \frac{1}{S-1} \sum \limits_{s \neq i}^S \frac{a_i}{a_s}  \left(  \sum_{k = 1}^{L} \sum_{m = 1}^{L} \phi_{skm}^{(2)} \Cov{}{F_k^{\{-i\}}}{F_m^{\{-i\}}} \right) \right]}_{\Delta N_i \colon \text{Relative nonlinearity}} \\
    + & \underbrace{\left(\sum_{k = 1}^{L} \zeta_{ik}^{(1)} \Cov{}{E_i}{F_k^{\{-i\}}} \right)   - \frac{1}{S-1} \sum \limits_{s \neq i}^S \frac{a_i}{a_s} \left( \sum_{k = 1}^{L} \zeta_{sk}^{(1)} \Cov{}{E_s}{F_k^{\{-i\}}} \right). }_{\Delta I_i \colon \text{The storage effect}}
\end{aligned}
\end{equation}

But why do we need to correct for speed? The simple answer is that species with fast population dynamics will dominate the invader--resident comparison. The more elaborate answer is that we want coexistence mechanisms to measure the degree to which specialization/differentiation contributes to coexistence, and differences in population-dynamical speed are not the sort of between-species differences that lead to coexistence, so they should not have a dramatic effect on the values of coexistence mechanisms. As a simple illustration of this, consider the two-species competitive Lotka-Volterra Model,
\begin{equation}
\label{LV1}
\frac{1}{n_j(t)} \frac{d n_j(t)}{d t} = b_j(k_j - \sum_{k = 1}^2 \alpha_{jk} n_{k}(t)), \quad j = (1,2).
\end{equation}
Using the typical invasion analysis, we find that coexistence is attained when $\alpha_{12} / \alpha_{22} < k_1/k_2 < \alpha_{11} / \alpha_{21}$. Note that the speed parameters, $b_j$, appear nowhere in the coexistence criterion. Though the speed parameters can modulate the invasion growth rates, what really matters is whether the invasion growth rates are positive or negative (\cite{Schreiber2011}; \cite{pande2020mean}).

To be slightly more general, consider a community with basic resource--consumer dynamics. If species do not specialize (i.e., they consume resources per capita in exactly the same proportions), then the average deviation from equilibrium resource levels, $\overline{F_k^{\{-i\}}} - F_k^{*j}$, will be the same regardless of which species is in the invader state (this is assuming that the $F_k^{*j}$ are the same for each species). Yet, when species have different population-dynamical speeds, their responses to fluctuations in regulating factors, $\phi_{jk}^{(1)}$, can be quite different. Large between-species differences in $\phi_{jk}^{(1)}$ can amplify the fluctuation $\overline{F_k^{\{-i\}}} - F_k^{*j}$ (which is generically non-zero in stochastic models due to nonlinear averaging), resulting in a substantially non-zero $\Delta \rho_i$ (see \eqref{MCT simple}). A substantially positive $\Delta \rho_i$ indicates that species $i$ is specializing on the mean levels of some regulating factor(s), even though species are not specialized at all. This undesirable behavior can be remediated by the speed conversion factors, which hypothetically give each species the same population-dynamical speed.

Population-dynamical speed is not always irrelevant to coexistence. For example, fast-growing species will dampen fluctuations in resource supply, thus modulating relative nonlinearity (\cite{hsu1980competition}; \cite{smith1981competitive}). Additionally, fast-growing species will quickly build up population size when the environment is favorable, thus modulating the storage effect (\cite{li2016effects}). However, we conjecture that slow-fast life history differences per se do not lead to coexistence. We do not know of any models in which species can coexist, despite being identical in all respects except for their population-dynamical speed. Even if such a model did exist, the speed conversion factors would not obscure this ostensibly-novel coexistence-promoting mechanism; they would simply re-scale growth rates to be more even.

The speed of population dynamics, $a_j$, does not have a precise definition, but it can often be understood in relation to the non-dimensionalization of a single-species model (\cite{Nisbet1982Modelling}, p. 21). For example, in the aforementioned Lotka-Volterra Model (\eqref{LV1}), the time variable can be factored as $t = \tau \times \hat{t}$, where $\tau$ is a dimensionless variable and $\hat{t}$ are the units. Selecting $\hat{t} = b_j$, the model can be re-written as  

\begin{equation}
\frac{1}{n_j(t)} \frac{d n_j(t)}{d \tau} = k_j - \sum_{k = 1}^2 \alpha_{jk} n_{k}(t).
\end{equation}

The characteristic time-scale of dynamics is $1/\hat{t} = 1/b_j$, so $1/b_j$ is the characteristic time-scale of population dynamics, or equivalently, $b_j$ is the characteristic rate of population dynamics. Here there is a critical problem: to understand coexistence in real communities, one must fit a model to data; but if one were to parameterize \eqref{LV1} with data, the $b_j$ will be statistically unidentifiable (\cite[p.~365]{gelman2014bayesian}). That is, in lieu of prior information about parameter values, an infinite number of parameter combinations will produce the same likelihood, leading to massive parameter uncertainty and instability in model-fitting routines. 

In general, population speed does not always present itself as a pre-existing parameter. What then, is a general-use definition of $a_j$? Our recommendation is to operationalize population-dynamical speed as the reciprocal of generation time (GT), i.e., $a_j = 1 / \text{GT}_j$. With this choice, the speed conversion factors become a ratio of generation times: $a_i/a_s = \text{GT}_s/\text{GT}_i$. While there is no single definition of generation time in structured population models (\cite[Section 5.3]{CaswellHal2001Mpm:}), many definitions are equivalent when species have attained their limiting dynamics (\cite{ellner2018generation}).

Species' generation times --- and thus the speed parameters --- are constants in many models (e.g., Case study \#2, Section \ref{Case studies}). However, it is possible for generation time to change depending on which species is in the invader state. In such models, there are two solutions: 1) calculate generation time in the context of the limiting dynamics of the full community (i.e., no invaders); or 2) calculate the generation time for each sub-community where a different species is in the invader state, in which case the notation $a_j^{\{-i\}}$ would become appropriate. We tentatively recommend the latter method. 

To calculate generation time,  one must have distinct information about mortality and reproduction. Unfortunately, some population models (particularly unstructured models) only include the aggregate effects of mortality and reproduction. For example, the term $\left(- \sum_{k = 1}^S \alpha_{jk} n_{k}(t)\right)$ in the Lotka-Volterra model (\eqref{LV1}) could represent the fact the birth rates decrease with population density or that death rates increase with population density (\cite[p.~125]{allen2010introduction}). 

In what follows, we present two alternative ways to operationalize \textit{speed} when generation time cannot be unambiguously derived from a model: 

\begin{enumerate}

    \item Obtain generation times from public databases (see \cite{de2009database}; \cite{jones2009pantheria}; \cite{myhrvold2015amniote}). The downside of this approach is that the speed conversion factors are fixed point estimates; because they do not vary with model parameters, there is a risk of underestimating the uncertainty in the values of coexistence mechanisms.
    
    \item Operationalize \textit{speed} as the sum of the magnitudes of sensitivities to the determinants of the average per capita growth rates: 

\begin{equation} \label{weird_speed}
    a_j = \abs{ \alpha_j^{(1)}} + \abs{ \alpha_j^{(2)}}  + \left(\sum_{k = 1}^L \abs{ \phi_{jk}^{(1)}} \right) + \left(\sum_{k = 1}^L \sum_{m = 1}^L \abs{ \phi_{jkm}^{(2)}} \right) + \left(\sum_{k = 1}^L \abs{ \zeta_{jk}^{(1)}} \right).
\end{equation}

The idea is that if a species responds strongly to the factors that control the average growth rate (e.g., mean fluctuations in regulating factors, covariances between the environmental parameter and regulating factors; see \eqref{taylor_decomp_avg1}, Section \ref{A summary of Modern Coexistence Theory}), then it likely has a fast life-cycles. This choice of $a_j$ has the benefit of being universally applicable: once a few analysis decisions have been made (i.e., $E_j$ and $\boldsymbol{F}$ are identified, the equilibrium parameters are selected), computing the speed parameters is trivial. However, with this approach, the speed parameters may be sensitive to model specification.

To illustrate this shortcoming, once again consider the competitive Lotka-Volterra model (\eqref{LV1}). The natural choice for the regulating factors are the species' densities, which means that the $\phi_{jk}^{(1)}$'s are the competition coefficients. \textcite{macarthur1970species} showed that the competitive Lotka-Volterra model is equivalent to a resource-consumer model in the limit of fast resource dynamics. In a special case of this resource-consumer model, the competition coefficients taken on a very simple form: $\alpha_{jk} = \sum_l^L c_{jl} c_{kl}$, where $c_{jl}$ is the per capita rate at which species $j$ consumes resource $l$. 

If one uses the Lotka-Volterra approximation of the resource-consumer model, then $a_j = \sum_{k = 1}^L \abs{ \alpha_{jk}}  = \sum_k^S \abs{\sum_l^L c_{jl} c_{kl}}$. However, if one was to study the more fundamental resource consumer model, then the regulating factors are the resources, $\phi_{jk}^{(1)} = c_{jk}$, and thus $a_j = \sum_{l = 1}^L \abs{c_{jl}}$. Clearly, the modeller's perspective and the level of mechanistic detail will influence the $a_j$ as defined by \eqref{weird_speed}. This is good to keep in mind, but it is by no means a knock-down argument against the use of \eqref{weird_speed}. After all, conclusions always depend (to some degree) on a researcher's perspective. Specifically in the context of MCT, it is well known that conclusion depend on the definitions of $E_j$ and $\boldsymbol{F}$, as well as their equilibrium values (\cite{barabas2018chesson}).

\end{enumerate}

\subsubsection{The relationship between scaling factors and speed conversion factors}
\label{The relationship between scaling factors and speed conversion factors}

In models with a single regulating factor, the scaling factors are
\begin{equation}
    q_{is} = \frac{\phi_{i1}^{(1)}}{\phi_{s1}^{(1)}}.
\end{equation}
This quotient resembles the speed conversion factors, $a_i/a_s$. As it turns out, the scaling factors and speed conversion factors are equivalent in certain models. In the lottery model of reef fish dynamics (a seminal model in coexistence theory; \cite{chesson1981environmentalST}), the finite rate of increase can be written as 
\begin{equation}
    \lambda_j = \exp{E_j - F_1} + (1-d_j),
\end{equation}
where $d_j$ is the death probability for adult fish. Species' sensitivity to $F_1$ (here, the logarithm of fish larvae per open territory) is $\phi_{j1}^{(1)} =  -d_j$. The scaling factors are therefore 

\begin{equation}
    q_{is} = \frac{d_i}{d_s}.
\end{equation}

There are several definitions of generation time (\cite{CaswellHal2001Mpm:}; \cite{bienvenu2015new}; \cite{ellner2018generation}), but when populations are at equilibrium, and when the state of offspring are independent of the state of parents, all definitions of generation time are equivalent to the mean age of the parents of offspring (\cite{ellner2018generation}). The aforementioned conditions hold true in the lottery model, and additionally, offspring production is independent of adult age, meaning that generation time is equal to the average adult lifespan. Because each individual fish has an independent and equal probability of death in each time-step, adult lifespan is distributed via a geometric distribution with mean $1/d_j$. Therefore, in the lottery model, the scaling factors are equivalent to the quotient of generation times. Or in other words, the scaling factors are equivalent to the speed conversion factors. The exact same equivalence also appears in the annual plant model (\cite{Chesson1994}), another seminal model in coexistence theory.

This equivalence is between speed conversion factors and scaling factors is incidental, but perfectly explicable: the scaling factors act as a conversion between species' sensitivities to regulating factors, and generation time will undoubtedly influence how quickly a species responds to changes in the regulating factors. That being said, the scaling factors do not convert between species' typical/average sensitivity to competition, which is what generation time modulates. Rather, the scaling factors convert between species' sensitivities to regulating factors, for each factor individually.

Historically, MCT has been used to analyze models with either a single regulating factor, or in models in which species interact equally with all heterospecifics (e.g., \cite{Chesson1994}; \cite{chesson2000mechanisms}), sometimes referred to as \textit{diffuse competition} (\cite{stump2017multispecies}). The simplifying structure of diffuse competition allows one to derive mathematical formulas in the $S$-species case, which is often otherwise intractable. In fact, when the regulating factors are species' densities, and the sensitivities $\boldsymbol{\Phi}$ are defined by a matrix of competition coefficients with diagonal elements $c$ (i.e., intraspecific competition) and off-diagonal elements $x$ (i.e., interspecific competition), we can derive (Appendix \ref{Scaling factors in the case of diffuse competition}) an exceedingly simple formula:

\begin{equation}
    q_{is} = \frac{x}{c + (S-2) x}.
\end{equation}

One problem with scaling factors is that they can become very large via a matrix inversion, and can therefore alter other coexistence mechanisms in a counterintuitive way (see case study \#1, Section \ref{Case studies}). The formulas in this section demonstrate why this problem is not readily apparent: the simplifying assumptions made by theoretical ecologists result in scaling factors that are reasonable. In the case of diffuse competition, the scaling factors become equivalent to the simple comparison as $S$ grows large (both converge to $1/S$). In the case of a single regulating factor, the scaling factors are equivalent to the speed conversion factors (at least in some simple models).

\subsection{The invader--invader comparison}
\label{The invader--invader comparison}

So far, we have been operating under the implicit assumption that coexistence mechanisms should be calculated as invader--resident comparisons. At first glance, this seems appropriate: Coexistence mechanisms are supposed to measure the importance of different explanations for coexistence, the concept of specialization/differentiation has played a central role in historical explanations of coexistence, and the invader--resident comparison putatively captures the notion of specialization/differentiation. However, upon further reflection, we may worry that the invader--resident comparison not only captures the rare-species advantage that results from specialization, but also that which results from intrinsic, density-independent differences between species. 

The alternative to the invader--resident comparison is what we call the \textit{invader--invader comparison}: a comparison of the high density and low density states of a single focal species. An invader--invader comparison holds species-specific features constant, thus isolating the effects of rarity. Using the \textit{difference-making / but-for account of causation} (\cite{sep-causation-law}), we can say that the invader--invader comparison gives the causal effects (on average per capita growth rates) of perturbing a species to low density, mediated through different variables (e.g., mean resource levels for $\Delta \rho_i$, resource variation for $\Delta N_i$).

The inventor of MCT, Peter Chesson, has previously alluded to the invader--invader comparison: "Often the mechanism is most easily understood in terms of how the conditions encountered by an individual species change between its resident and invader states." (\cite{Chesson2008}), and "...within-species comparison is more reliable if appropriate within-species resident and invaders states can be prepared" (\cite{chesson2013species}). To our knowledge, no papers have attempted to use the invader--invader versions of coexistence mechanisms, which we define in the following partition of the invasion growth rate:

\begin{equation} \label{MCT invader}
\begin{aligned}
    \overline{r}_i \approx & \underbrace{ \left( \sum_{k = 1}^{L} \phi_{ik}^{(1)} \left(\overline{F_k^{\{-i\}}} - F_k^{*i} \right) \right) -   \left( \sum_{k = 1}^{L} \phi_{ik}^{(1)} \left(\overline{F_k} - F_k^{*i} \right) \right) }_{\Delta \rho_i \colon \text{Linear density-dependent effects}} \\
    + & \underbrace{\frac{1}{2}\left[ \left( \sum_{k = 1}^{L} \sum_{m = 1}^{L} \phi_{ikm}^{(2)} \Cov{}{F_k^{\{-i\}}}{F_m^{\{-i\}}} \right)  -  \left(  \sum_{k = 1}^{L} \sum_{m = 1}^{L} \phi_{ikm}^{(2)} \Cov{}{F_k}{F_m} \right) \right]}_{\Delta N_i \colon \text{Relative nonlinearity}} \\
    + & \underbrace{\left(\sum_{k = 1}^{L} \zeta_{ik}^{(1)} \Cov{}{E_i}{F_k^{\{-i\}}} \right)   -  \left( \sum_{k = 1}^{L} \zeta_{ik}^{(1)} \Cov{}{E_s}{F_k} \right), }_{\Delta I_i \colon \text{The storage effect}}
\end{aligned}
\end{equation}

Note the absence of the superscript "$\{-i\}$" from the subtracted terms, which indicates that the reference state is a community where all species are at their typical abundances. Also note that the density-independent effects, $\Delta E_i$, have vanished.

Unfortunately, the invader--invader comparison lacks the generality of the invader--resident comparison. There may be no stable high-density state for the focal species, as is the case when the focal species has a negative invasion growth rate, or when the focal species becomes temporarily abundant only to become excluded later on (a phenomenon that has been dubbed \textit{the resident strikes back}; \cite{Mylius2001TheAttractor}; \cite{Geritz2002}). When the invader--invader comparison does exist, it will not be unique if there are multiple stable high-density states for the focal species. In our opinion, the most troubling problem with the invader--invader comparison is that it cannot cover the case of negative invasion growth rates; this prevents us from learning about how species are failing to coexist. 

The invader-invader comparison is successful, insofar as it does not included intri

It is worth belaboring the distinction between the biological interpretations of the invader--resident comparison and the invader--invader comparison. In an invader--resident coexistence mechanism, differences between invaders and residents can be caused by both 1) specialization that creates a rare-species advantage, and 2) intrinsic differences between species that are not part of a density-dependent feedback loop. By contrast, the invader--invader coexistence mechanisms isolates the effects of rarity, but does not always capture the notion of specialization/differentiation. It is true that some form of specialization/differentiation is needed for a positive invasion growth rate (\cite{chesson1991need}), and therefore, that invader--invader coexistence mechanisms represent specialization in some ultimate sense. However, because different species are not directly compared, different form of specialization are distributed to invader-invader coexistence mechanisms in a non-obvious fashion. For example, in case study \#1 (Section \ref{Case studies}), we show that the invader--invader version of relative nonlinearity can be large and positive, despite the fact that the focal species specializes on means resource levels, not resource variance. The lesson here is not that invader--invader comparisons are faulty, but that they should not naively interpreted as capturing \textit{specialization}.
 
\section{Case studies}
\label{Case studies}

Modern Coexistence Theory (MCT) systematizes the analysis of models by breaking an arbitrarily complex growth rate function into simple polynomial terms, and therefore, is most useful when we don't know how species are coexisting. However, if MCT is to be a useful measurement tool, it ought to give us answers that make sense in models where we do know how species are coexisting.

In this section, we analyze two models in which it is intuitively clear how species are coexisting and then check to see which methods agree with intuition. Both models describe the dynamics of three species, since four-species models would be unnecessarily complicated, and two-species models would not reveal the pathological behavior of scaling factors (See Section \ref{The relationship between scaling factors and speed conversion factors}). For each model, we examine two sets of parameter values: one where species have similar population-dynamical speeds, and one where a single species grows much faster than others. All computations can be replicated using the Mathematica notebooks, {\fontfamily{qcr}\selectfont ArmMc\_3Spp.nb} and  {\fontfamily{qcr}\selectfont SE\_3Spp.nb}, found at \url{https://github.com/ejohnson6767/scaling_factors}.

\subsection{Case study \#1: Coexistence via relative nonlinearity in a resource-consumer model}
\label{Coexistence via relative nonlinearity in a resource-consumer model}

Here, we examine a deterministic, continuous-time resource-consumer model, inspired by the Armstrong-McGehee model (\citeyear{armstrong1976coexistence}; \citeyear{armstrong1980competitive}). The equations for three consumers (densities denoted by $N_1$, $N_2$, and $N_3$) and two resources (densities denoted by $R_1$ and $R_2$) are

\begin{equation}
\frac{d N_1}{d t} = N_1  b_1 \left[ c_{11} R_1 + c_{21} R_2 - d \right]    
\end{equation}

\begin{equation}
\frac{d N_2}{d t} = N_2   b_2 \left[ c_{12} R_1 + \frac{ c_{22} R_2}{\eta + R_2} - d \right]       
\end{equation}

\begin{equation}
\frac{d N_3}{d t} = N_3   b_3 \left[ c_{13} R_1 + c_{23} R_2 - d \right]      
\end{equation}

\begin{equation}
\frac{d R_1}{d t} = R_1 \left[ r_1 \left( 1 - \frac{R_1}{K_1} \right) - c_{11} N_1 - c_{12} N_2 - c_{13} N_3 \right]
\end{equation}

\begin{equation}
\frac{d R_2}{d t} = R_2 \left[ r_2 \left( 1 - \frac{R_2}{K_2} \right) - c_{21} N_1 - \frac{ c_{22} N_2}{\eta + R_2} - c_{23} N_3 \right].
\end{equation}

In the absence of consumers, both resources grow logistically  with intrinsic growth rates $r_j$ and carrying capacities $K_j$. Consumers die at a shared density-independent rate, $d$, and have birth rates proportional to resource consumption. The maximum (per-consumer, per-resource) rate at which resource $k$ is consumed by consumer $j$ is given by $c_{kj}$.  Consumers 1 and 3 have linear functional responses to resource densities. Consumer 2 has a linear response to resource 1, but has a type II functional response to resource 2 with a half-saturation constant $\eta$. Population-dynamical speed is denoted by $b_j$. 

We selected parameter values so that consumer 1 specializes on resource 1, consumer 2 specializes on resource 2 and consumer 3 specializes on the variation in resource 2: $c_{11} = c_{22} = c_{23} = 1, \; c_{21} = c_{12} = c_{13} = 0.05 \; d_1 = d_2 = d_3 = 0.47, \; r_1 = r_2 = 1, \; K_1 = K_2 = 1.5, \; \eta = 0.5$. These parameter values produce two virtually independent subsystems: \{consumer 1, resource 1\} and \{consumer 2, consumer 3, resource 2\}, the latter of which is essentially the Armstrong-McGehee model. The regulating factors are simply the resource densities. The environmental parameter $E_j$ is nonexistent, so $\Delta E_i$ and $\Delta I_i$ are necessarily zero. 

\begin{figure}[h]
 \centering
      \includegraphics[scale = 0.5]{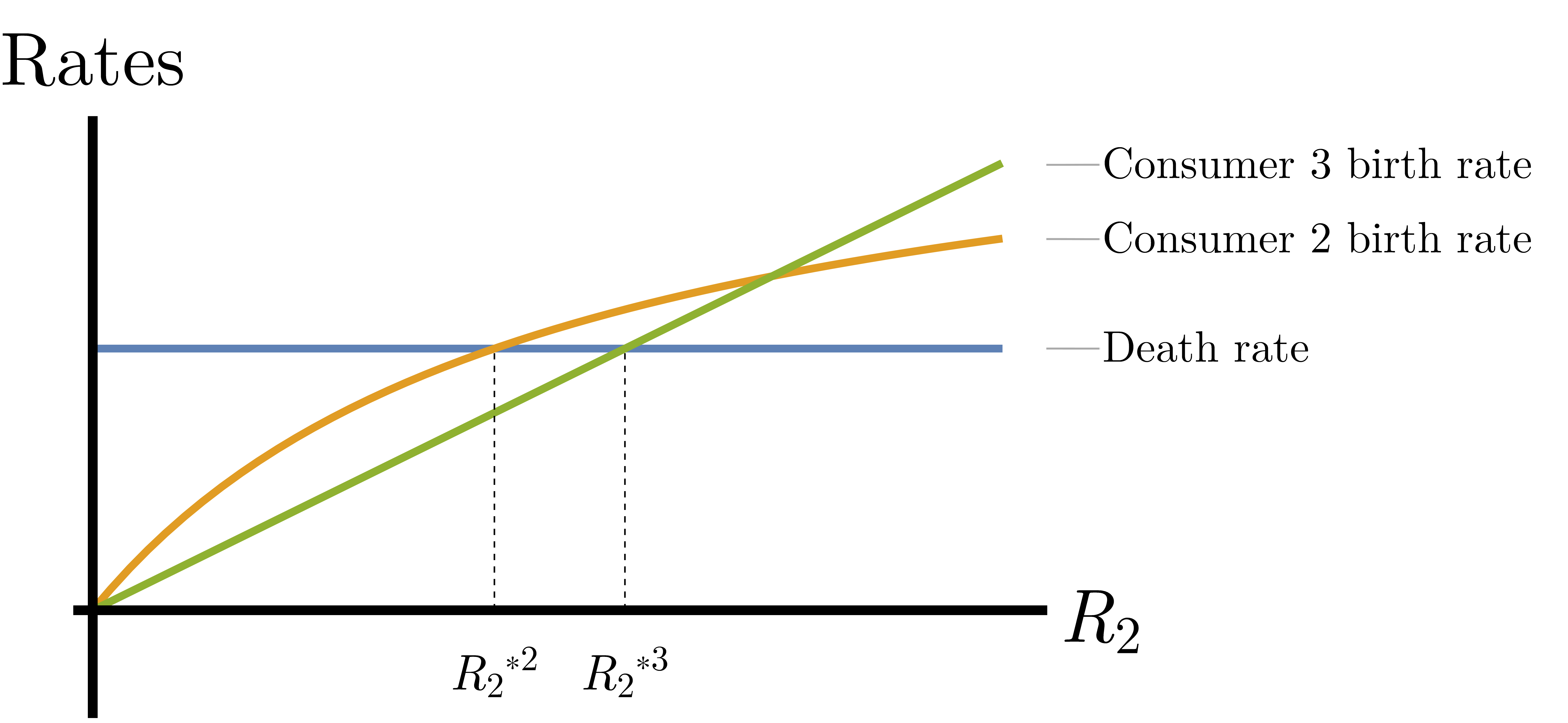}
  \caption{An opportunist-gleaner trade-off. Consumer 2 (the \textit{gleaner}) excludes consumer 3 (the \textit{opportunist}) in the absence of resource fluctuations (by \posscite{tilman1982resourceST} $R^*$ rule: $R_2^* < R_3^*$), but consumer 2 is hurt more by resource fluctuations (by \posscite{jensen1906fonctions} inequality). Consumer 2 specializes on mean resource levels, whereas Consumer 3 specializes on resource variation. Consumer 1 is not shown.}
      \label{fig:Opportunist Gleaner}
\end{figure}

Consumer 1 specializes on resource 1, and thus coexists via the linear density-dependent effects. Because consumer 2 and consumer 3 both heavily consume the same resource, one of these species must coexist via fluctuation-dependent mechanisms. Consumer 2 clearly coexists via linear density-dependent effects, because it is the superior competitor (compared to consumer 3) in the absence of fluctuations via Tilman's $R^*$ rule (see Fig. \ref{fig:Opportunist Gleaner}). Consumer 3 clearly coexists via relative nonlinearity, because consumer 2's birth rate function is relatively concave down, meaning that resource fluctuations help consumer 3 relative to consumer 2.

Following the intuition in the previous paragraph, we predict that $\Delta \rho_1$, $\Delta \rho_2$, and $\Delta N_3$ will be positive and large (relative to other coexistence mechanisms within each species, respectively). This is precisely what we see for both the simple comparison and the speed conversion factors (Table \ref{AM_same_speed}). By contrast, the scaling factors counterintuitively attribute the persistence of species 1 to relative nonlinearity (i.e., $\Delta N_1$ is large and positive). It is not so surprising that $\Delta N_1$ is non-zero: after all, species 1 does have a nonlinear response to competition, relative to species 2. What is surprising is that $\Delta N_1$ is so large that it almost entirely accounts for the positive invasion growth rate of species 1. Of course, it is unreasonable to think that $\Delta \rho_1$ would be large and positive when using the scaling factors (the whole point of the scaling factors is to cancel $\Delta \rho_i$), but one might reasonably think that $r'_1$ would be large and positive, since both $\Delta \rho_1$ and $r'_1$ represent fluctuation-independent forces.

The failure of the scaling factors method (i.e., the counterintuitively large $\Delta N_1$) can be explained by the sheer magnitude of the scaling factors, which are $q_{12} \approx -84$ and $q_{13} \approx 64$. Because species 2 barely interacts with species 1, the growth rate components of species 2 must be heavily weighted in order to cancel $\Delta \rho_1$. Consider the analytical formula for one scaling factor, $q_{12} = (\phi_{12}^{(1)} \phi_{31}^{(1)} - \phi_{11}^{(1)} \phi_{32}^{(1)}) / (\phi_{22}^{(1)} \phi_{31}^{(1)} - \phi_{21}^{(1)} \phi_{32}^{(1)})$. This formula demonstrates that the scaling factors can become large via division by a small number; Because $\phi_{31}$ and $\phi_{21}$ are small, the denominator $\phi_{22}^{(1)} \phi_{31}^{(1)} - \phi_{21}^{(1)} \phi_{32}^{(1)}$ is small, even though the species respond similarly to resources in total. 

When species have similar population-dynamical speeds, the simple comparison and speed conversion factors both give results that accord with intuition (Table \ref{AM_same_speed}). Strangely, the invader--invader comparison produces exactly the opposite of what we predicted for species 2 and 3: the coexistence mechanisms $\Delta N_2$ and  $\Delta \rho_3$ are large and positive.

How can we make sense of the invader--invader coexistence mechanisms? In our model, consumer 2 and consumer 3 exhibit an \textit{opportunist-gleaner trade-off} (Fig. \ref{fig:Opportunist Gleaner}; \cite{grover1997resource}). When species 2 --- the gleaner --- becomes abundant, it increases resource variation by inducing cyclical resource-consumer dynamics. Because the gleaner has a concave-down per capita growth rate function, its high-density state suffers from the increased variation, resulting in $\Delta N_2 > 0$. When the opportunist --- species 3 --- is absent from the community, the gleaner produces a lower $\overline{F_2^{\{-3\}}}$ through nonlinear averaging. When the opportunist is at its high-density state, resource fluctuations becomes smaller, nonlinear averaging becomes weaker, and mean resource level rise, resulting in $\Delta \rho_3 > 0$. Here we have shown that is possible to make sense of the invader--invader coexistence mechanisms; that is to say, the counterintuitive values of the invader--invader coexistence mechanisms is not necessarily a sign of failure, but rather a reflection of our desire to think of coexistence mechanisms as measuring \textit{specialization}. 

To better understand the effects of population-dynamical speed on the coexistence mechanisms, we increase species $1$'s speed by setting $b_1 = 100$. Because species 1 attains an equilibrium with resource 1 and barely interacts with the subsystem \{consumer 2, consumer 3, resource 2\}, increasing the speed of species 1's population dynamics has little effect on population dynamics of the full three-species community, or any of the sub-communities (see simulated time series in the Mathematica notebook {\fontfamily{qcr}\selectfont ArmMc\_3Spp.nb}.

Table \ref{AM_diff_speed} shows that increasing species 1's speed can complicate the interpretation of species 3's coexistence mechanisms. Species 3 specializes on resource variation relative to species 2, so $\Delta N_3$ should be large; additionally, species 3 specializes on resource 2 relative to species 1, so perhaps $\Delta \rho_{3,F_2}$ will be substantially positive. This is what we see for the speed conversion factor method, but the simple comparison method gives us a large $\Delta \rho_{3,F_1}$, implying that species 3 persists by specializing on resource 1.

Because species 3 barely consumes resource 1, the mean level of resource 1 barely changes species 3 is perturbed to the invader state. However, there is a small persistent difference between $\overline{F_1^{\{-3\}}}$ and $F_1^{*1}$, not because of species 3's interaction with resource 1, but because the community is a nonlinear and non-equilibrium system (because of nonlinear averaging, it is not possible to select $\boldsymbol{F^{*1}} = \overline{\boldsymbol{F}^{\{-3\}}}$ and still satisfy the constraint $g_1(E_1^*, \boldsymbol{F^{*1}}) = 0$). The small difference between $\overline{F_1^{\{-3\}}}$ and $F_1^{*1}$ gets amplified by species 1's extreme responsiveness to regulating factors (i.e., a large $\phi_{11}^{(1)}$), which can be attributed to the species $1$'s fast population dynamics. The term $\phi_{11}^{(1)} \left(\overline{F_1} - F_1^{*1} \right)$, belonging to species $1$, comes to dominate in the simple comparison method, resulting in a large $\Delta \rho_{3,F_1}$. The speed conversion factors method successfully counteracts this phenomenon.

\newpage

\begin{table}[H]
\label{AM_same_speed}
\caption{Case study \#1: Values of coexistence mechanisms in the scenario where all species have the same population-dynamical speed. The superscript "$(e)$" denotes exact coexistence mechanisms. The density-independent effects are denoted by $r'_i$ (only for the scaling factor method) or $\Delta E_i$. The \textit{subjective} version of the speed conversion factors were obtained by selecting $a_j = b_j$; The \textit{universal} version of the the speed conversion factors were obtained via \eqref{weird_speed} in Section \ref{Speed conversion factors}.}
\tiny
\hspace*{-2cm}
\begin{tabular}{p{0.055 \linewidth} l p{0.045 \linewidth}p{0.045 \linewidth} p{0.045 \linewidth}p{0.045 \linewidth}p{0.045 \linewidth}p{0.045 \linewidth} p{0.045 \linewidth}p{0.045 \linewidth}p{0.045 \linewidth}p{0.045 \linewidth}p{0.065 \linewidth}p{0.045 \linewidth} }
\toprule \\
\textbf{Species} $i\,=\,\ldots$ & \multicolumn{1}{c}{\textbf{Calculation method}} & \multicolumn{10}{c}{\textbf{Coexistence mechanisms}} & \multicolumn{2}{c}{\textbf{Invasion growth rate}}\\
\\
& & \multicolumn{5}{c}{small-noise} & \multicolumn{5}{c}{exact} & \multicolumn{1}{c}{approx} & \multicolumn{1}{c}{exact} \\ 
\cmidrule(lr){3-7} \cmidrule(lr){8-12} \cmidrule(lr){13-13} \cmidrule(lr){14-14} \\
 & & $r_i^\prime$or $\Delta E_i$ & $\Delta \rho_i$ & $\Delta \rho_{i,F_{1}}$ & $\Delta \rho_{i,F_{2}}$ & $\Delta N_i$ & $r_{i}^{\prime (e)}$or $\Delta E_i^{(e)}$ & $\Delta \rho_{i}^{(e)}$ & $\Delta \rho_{i,F_{1}}^{(e)}$ & $\Delta \rho_{i,F_{2}}^{(e)}$ & $\Delta N_{i}^{(e)}$ & $\approx \overline{r_i^{\{-i\}}}$ & $\overline{r_i^{\{-i\}}}$ \\ 
\midrule \\
 1 & \text{Scaling factors} & -7.922 & 0 & \text{NA} & \text{NA} & 10.855 & -7.498 & 0 & \text{NA} & \text{NA} & 8.520 & 2.932 &
   1.010 \\
 1 & \text{Simple comparison} & 0 & 0.957 & 0.962 & -0.005 & 0.065 & 0 & 0.960 & 0.962 & 0.048 & 0.051 & 1.022 & 1.010 \\
 1 & \text{Speed conversion factors (subjective)} & 0 & 0.957 & 0.962 & -0.005 & 0.065 & 0 & 0.960 & 0.962 & 0.048 & 0.051 &
   1.022 & 1.010 \\
 1 & \text{Speed conversion factors (universal)} & 0 & 0.991 & 0.978 & 0.013 & 0.023 & 0 & 0.992 & 0.978 & 0.032 & 0.018 & 1.015
   & 1.010 \\
 1 & \text{Invader--invader comparison} & 0 & 1.010 & 1.013 & -0.002 & 0.000 & 0 & 1.010 & 1.013 & -0.002 & -0.000 & 1.010 &
   1.010 \\
   \midrule \\
 2 & \text{Scaling factors} & 0.094 & 0 & \text{NA} & \text{NA} & 0.000 & 0.080 & 0 & \text{NA} & \text{NA} & 0.000 & 0.094 &
   0.080 \\
 2 & \text{Simple comparison} & 0 & 0.094 & -0.000 & 0.094 & 0.000 & 0 & 0.080 & -0.000 & 0.080 & 0.000 & 0.094 & 0.080 \\
 2 & \text{Speed conversion factors (subjective)} & 0 & 0.094 & -0.000 & 0.094 & 0.000 & 0 & 0.080 & -0.000 & 0.080 & 0.000 &
   0.094 & 0.080 \\
 2 & \text{Speed conversion factors (universal)} & 0 & 0.094 & -0.000 & 0.094 & 0.000 & 0 & 0.080 & -0.000 & 0.081 & 0.000 &
   0.094 & 0.080 \\
 2 & \text{Invader--invader comparison} & 0 & 0.001 & -0.000 & 0.001 & 0.117 & 0 & 0.001 & -0.000 & 0.082 & 0.081 & 0.118 & 0.080
   \\
   \midrule \\
3 & \text{Scaling factors} & -0.124 & 0 & \text{NA} & \text{NA} & 0.228 & -0.088 & 0 & \text{NA} & \text{NA} & 0.141 & 0.104 &
   0.054 \\
 3 & \text{Simple comparison} & 0 & -0.013 & 0.001 & -0.015 & 0.087 & 0 & -0.000 & 0.001 & 0.052 & 0.054 & 0.073 & 0.054 \\
 3 & \text{Speed conversion factors (subjective)} & 0 & -0.013 & 0.001 & -0.015 & 0.087 & 0 & -0.000 & 0.001 & 0.052 & 0.054 &
   0.073 & 0.054 \\
 3 & \text{Speed conversion factors (universal)} & 0 & -0.013 & 0.000 & -0.013 & 0.087 & 0 & -0.000 & 0.000 & 0.054 & 0.054 &
   0.073 & 0.054 \\
 3 & \text{Invader--invader comparison} & 0 & 0.056 & -0.000 & 0.056 & 0.000 & 0 & 0.056 & -0.000 & 0.056 & -0.000 & 0.056 &
   0.054 \\
\bottomrule \\ 
\end{tabular}
\hspace*{-2cm}
\end{table}

\begin{table}[H]
\label{AM_diff_speed}
\caption{Case study \#1: Values of coexistence mechanisms in the scenario where species 1 has much faster population dynamics than species 2 and 3.}
\tiny
\hspace*{-2cm}
\begin{tabular}{p{0.055 \linewidth} l p{0.045 \linewidth}p{0.045 \linewidth} p{0.045 \linewidth}p{0.045 \linewidth}p{0.045 \linewidth}p{0.045 \linewidth} p{0.045 \linewidth}p{0.045 \linewidth}p{0.045 \linewidth}p{0.045 \linewidth}p{0.065 \linewidth}p{0.045 \linewidth} }
\toprule \\
\textbf{Species} $i\,=\,\ldots$ & \multicolumn{1}{c}{\textbf{Calculation method}} & \multicolumn{10}{c}{\textbf{Coexistence mechanisms}} & \multicolumn{2}{c}{\textbf{Invasion growth rate}}\\
\\
& & \multicolumn{5}{c}{small-noise} & \multicolumn{5}{c}{exact} & \multicolumn{1}{c}{approx} & \multicolumn{1}{c}{exact} \\ 
\cmidrule(lr){3-7} \cmidrule(lr){8-12} \cmidrule(lr){13-13} \cmidrule(lr){14-14} \\
 & & $r_i^\prime$or $\Delta E_i$ & $\Delta \rho_i$ & $\Delta \rho_{i,F_{1}}$ & $\Delta \rho_{i,F_{2}}$ & $\Delta N_i$ & $r_{i}^{\prime (e)}$or $\Delta E_i^{(e)}$ & $\Delta \rho_{i}^{(e)}$ & $\Delta \rho_{i,F_{1}}^{(e)}$ & $\Delta \rho_{i,F_{2}}^{(e)}$ & $\Delta N_{i}^{(e)}$ & $\approx \overline{r_i^{\{-i\}}}$ & $\overline{r_i^{\{-i\}}}$ \\ 
\midrule \\
 1 & \text{Scaling factors} & -792.237 & 0 & \text{NA} & \text{NA} & 1085.440 & -749.826 & 0 & \text{NA} & \text{NA} & 851.984 & 293.206 & 101.029 \\
 1 & \text{Simple comparison} & 0 & 100.976 & 101.237 & -0.261 & 0.065 & 0 & 100.979 & 101.237 & -0.207 & 0.051 & 101.041 & 101.029 \\
 1 & \text{Speed conversion factors (subjective)} & 0 & 95.742 & 96.223 & -0.481 & 6.453 & 0 & 95.995 & 96.223 & 4.836 & 5.065 & 102.195 & 101.029 \\
 1 & \text{Speed conversion factors (universal)} & 0 & 99.136 & 97.845 & 1.292 & 2.321 & 0 & 99.227 & 97.845 & 3.204 & 1.821 & 101.457 & 101.029 \\
 1 & \text{Invader--invader comparison} & 0 & 101.029 & 101.288 & -0.259 & 0.000 & 0 & 101.029 & 101.288 & -0.259 & -0.000 & 101.029 & 101.029 \\
   \midrule \\
 2 & \text{Scaling factors} & 0.094 & 0 & \text{NA} & \text{NA} & 0.000 & 0.080 & 0 & \text{NA} & \text{NA} & 0.000 & 0.094 & 0.080 \\
 2 & \text{Simple comparison} & 0 & 0.094 & -0.002 & 0.096 & 0.000 & 0 & 0.080 & -0.002 & 0.082 & 0.000 & 0.094 & 0.080 \\
 2 & \text{Speed conversion factors (subjective)} & 0 & 0.094 & -0.000 & 0.094 & 0.000 & 0 & 0.080 & -0.000 & 0.080 & 0.000 & 0.094 & 0.080 \\
 2 & \text{Speed conversion factors (universal)} & 0 & 0.094 & -0.000 & 0.094 & 0.000 & 0 & 0.080 & -0.000 & 0.081 & 0.000 & 0.094 & 0.080 \\
 2 & \text{Invader--invader comparison} & 0 & -0.001 & 0.000 & -0.001 & 0.117 & 0 & -0.000 & 0.000 & 0.080 & 0.080 & 0.117 & 0.080 \\
   \midrule \\
 3 & \text{Scaling factors} & -0.124 & 0 & \text{NA} & \text{NA} & 0.234 & -0.088 & 0 & \text{NA} & \text{NA} & 0.144 & 0.110 & 0.056 \\
 3 & \text{Simple comparison} & 0 & -0.013 & 0.138 & -0.151 & 0.089 & 0 & 0.001 & 0.138 & -0.082 & 0.055 & 0.077 & 0.056 \\
 3 & \text{Speed conversion factors (subjective)} & 0 & -0.013 & 0.001 & -0.014 & 0.089 & 0 & 0.001 & 0.001 & 0.055 & 0.055 & 0.077 & 0.056 \\
 3 & \text{Speed conversion factors (universal)} & 0 & -0.013 & 0.007 & -0.019 & 0.089 & 0 & 0.001 & 0.007 & 0.049 & 0.055 & 0.077 & 0.056 \\
 3 & \text{Invader--invader comparison} & 0 & 0.055 & -0.000 & 0.055 & 0.000 & 0 & 0.055 & -0.000 & 0.055 & 0.000 & 0.055 & 0.056 \\
\bottomrule \\ 
\end{tabular}
\hspace*{-2cm}
\end{table}

\newpage

\subsection{Case study \#2: Coexistence via the storage effect in a temporally autocorrelated environment}
\label{Coexistence via the storage effect in a temporally autocorrelated environment}

Here, we examine a stochastic, discrete-time model, inspired by \textcite{li2016effects} and \textcite{schreiber2021positively}. Three competitors interact via Lotka-Volterra-like dynamics, and per capita fecundity fluctuates with environmental conditions. More specifically, maximum fecundity is temporally autocorrelated, which serves to generate the covariance between environment and competition that is needed for the storage effect (\cite{li2016effects}, \cite{letten2018species}; \cite{schreiber2021positively}). Through inter-generational population growth, a good environment in the present (for a common species) generically leads to high competition in the future. However, a correlation between the future environment and future competition can only be established if the future environment is similar to the present environment, i.e., if the environment is autocorrelated.  

The equations for the three consumers (densities denoted by $N_1$, $N_2$, and $N_3$) are

\begin{equation}
N_j(t+1) = N_j(t) \left[ s_j + \frac{\exp{E_j(t)}}{1 + \alpha_{j1} N_1(t) + \alpha_{j2} N_2(t) + \alpha_{j3} N_3(t)}\right], \quad j = (1,2,3),  
\end{equation}

where $s_j$ is the survival probability, $E_j$ (the environmental parameter) is the logarithm of maximum per capita fecundity, and the $\alpha_{jk}$ are the competition coefficients. Because we have distinct information about survival and fecundity, we can easily operationalize the speed parameters using generation time: $a_j = 1/(1-s_j)$. 

The environmental parameters follow autoregressive order-1 dynamics:

\begin{equation}
E_j(t+1) = \theta_j E_j(t) + \sigma_j \sqrt{1-\theta_j^2}  \epsilon_j(t), \quad \epsilon_j(t) \sim Normal(0,1), \quad j = (1,2,3).
\end{equation}

Here, the factor $\sqrt{1-\theta_j^2}$ ensures that the variance of the asymptotic stationary distribution of $E_j(t)$ is always $\sigma_j^2$, regardless of the value of the autoregressive parameter, $\theta_j$.

The regulating factors in this model are the species densities. Thus, when a species is placed in the invader state, one of the regulating factors vanishes, and the scaling factors can be used to cancel $\Delta \rho_i$. To represent the fact that a regulating factor has vanished, we invert the matrix of responses with the $i$-th row and $i$-th column removed (\cite[p.~250]{Chesson1994}), as opposed to only removing the row (as in \eqref{qir_sol}, Section \ref{Scaling factors}). 

We set the model parameters so that species 2 and 3 have temporally autocorrelated environmental parameters, but species 1 does not: $\theta_1 = 0$, $\theta_2 = \theta_3 = 0.95$. All species experience the same degree of environmental variation: $\sigma_1 = \sigma_2 =\sigma_3 = 0.5$. The matrix of competition coefficients is 

\begin{equation}
\boldsymbol{\alpha} = \begin{bmatrix}
1 & 0.01 & 0.01 \\
0.01 & 1 & 1.01 \\
0.01 & 1.01 & 1 \end{bmatrix}.
\end{equation}

With competition structured in this way, species 1 is nearly-independent of species 2 and 3. In the absence of fluctuations, the either species 2 or 3 is competitively excluded depending on initial conditions; However, these \textit{priority effects} may be overcome by the storage effect. Intuition dictates that $\Delta \rho_1$ ($\Delta r'_i$ in the case of scaling factors), $\Delta I_2$, and $\Delta I_3$ will be positive and large (relative to other coexistence mechanisms within each species). 

In the case where all species have the same population-dynamical speed (i.e., $s_1 = s_2 = s_3 = 0.9$), all methods agree with intuition (Table \ref{SE_same_speed}). The scaling factors work here because species 2 and 3 affect species 1 identically; in the invader--resident comparison, a large and positive $q_{12}$ is cancelled-out by a large and negative $q_{13}$. A careful reader may notice that in addition to the storage effect, relative nonlinearity appears to help species 2 and 3 (Table \ref{SE_same_speed}); this result is expected when survival rates vary across species (\cite{chesson2003quantifying}; \cite{Yuan2015}). 

In the case where species 1 has much faster population dynamics (i.e., $s_1 = 0.5$, $s_2 = s_3 = 0.9$), the simple comparison incorrectly implies that species 2 and 3 are persisting due to the linear effects of competition (Table \ref{SE_diff_speed}). The explanation of this result mirrors our explanation in case study \#1: For species 1, the temporal average of competition is slightly different from the equilibrium competition (due to the non-equilibrium and nonlinear nature of the system), and this small fluctuation gets amplified by species 1's fast population dynamics. Species 1 is unduly emphasized in the invader--resident comparison, leading to nonsensical values of coexistence mechanisms.

\newpage

\begin{table}[H]
\label{SE_same_speed}
\caption{Case study \#2: Values of coexistence mechanisms in the scenario where all species have the same population-dynamical speed. Species 2 and 3 behave symmetrically, so they have the same coexistence mechanisms. The superscript "$(e)$" denotes exact coexistence mechanisms. The density-independent effects are denoted by $r'_i$ (only for the scaling factor method) or $\Delta E_i$. The \textit{generation} version of the speed conversion factors were obtained by selecting $a_j = 1/(1-s_j)$; The \textit{universal} version of the the speed conversion factors were obtained via \eqref{weird_speed} in Section \ref{Speed conversion factors}.}
\tiny
\hspace*{-2cm}
\begin{tabular}[t]{p{0.055 \linewidth} l  l p{0.055 \linewidth}p{0.055 \linewidth}p{0.055 \linewidth}p{0.055 \linewidth}p{0.05 \linewidth}p{0.055 \linewidth}p{0.055 \linewidth}p{0.055 \linewidth}p{0.07 \linewidth} p{0.055 \linewidth} }
\toprule \\
\textbf{Species} $i\,=\,\ldots$ & \multicolumn{1}{c}{\textbf{Calculation method}} & \multicolumn{8}{c}{\textbf{Coexistence mechanisms}} & \multicolumn{2}{c}{\textbf{Invasion growth rate}}\\
& & \multicolumn{4}{c}{small-noise} & \multicolumn{4}{c}{exact} & \multicolumn{1}{c}{approx} & \multicolumn{1}{c}{exact} \\ 
\cmidrule(lr){3-6} \cmidrule(lr){7-10} \cmidrule(lr){11-11} \cmidrule(lr){12-12} \\
 & & $r_i^\prime$or $\Delta E_i$ & $\Delta \rho_i$ & $\Delta N_i$ & $\Delta I_i$ & $r_{i}^{\prime (e)}$or $\Delta E_i^{(e)}$ & $\Delta \rho_{i}^{(e)}$ & $\Delta N_{i}^{(e)}$ & $\Delta I_{i}^{(e)}$ & $\approx \overline{r_i^{\{-i\}}}$ & $\overline{r_i^{\{-i\}}}$ \\ 
\midrule \\
 1 & \text{Scaling factors} & 0.100 & 0.000 & -0.000 & 0.000 & 0.600 & 0 & 0.001 & 0.019 &
   0.100 & 0.620 \\
 1 & \text{Simple comparison} & 0.000 & 0.108 & -0.023 & 0.004 & 0.000 & 0.605 & -0.012 &
   0.026 & 0.090 & 0.620 \\
 1 & \text{Speed conversion factors (generation)} & 0.000 & 0.108 & -0.023 & 0.004 & 0.000
   & 0.605 & -0.012 & 0.026 & 0.090 & 0.620 \\
 1 & \text{Speed conversion factors (universal)} & 0.000 & 0.108 & -0.023 & 0.004 & 0.000
   & 0.605 & -0.012 & 0.026 & 0.090 & 0.620 \\
 1 & \text{Invader--invader comparison} & 0 & 0.102 & -0.002 & 0.000 & 0.000 & 0.600 &
   -0.000 & 0.020 & 0.100 & 0.620 \\
   \midrule \\
 \text{2 $\&$ 3} & \text{Scaling factors} & -0.001 & 0.000 & 0.000 & 0.008 & -0.001 & 0 &
   -0.000 & 0.014 & 0.007 & 0.013 \\
 \text{2 $\&$ 3} & \text{Simple comparison} & -0.000 & -0.001 & 0.011 & 0.004 & -0.000 &
   -0.001 & 0.007 & 0.008 & 0.013 & 0.013 \\
 \text{2 $\&$ 3} & \text{Speed conversion factors (generation)} & -0.000 & -0.001 & 0.011
   & 0.004 & -0.000 & -0.001 & 0.007 & 0.008 & 0.013 & 0.013 \\
 \text{2 $\&$ 3} & \text{Speed conversion factors (universal)} & -0.000 & -0.001 & 0.011 &
   0.004 & -0.000 & -0.001 & 0.007 & 0.008 & 0.013 & 0.013 \\
 \text{2 $\&$ 3} & \text{Invader--invader comparison} & 0 & 0.005 & 0.001 & 0.005 & -0.001
   & 0.003 & 0.002 & 0.008 & 0.010 & 0.013 \\
\bottomrule \\ 
\end{tabular}
\hspace*{-2cm}
\end{table}

\begin{table}[H]
\label{SE_diff_speed}
\caption{Case study \#2: Values of coexistence mechanisms in the scenario where species 1 has much faster population dynamics than species 2 and 3.}
\tiny
\hspace*{-2cm}
\begin{tabular}[t]{p{0.055 \linewidth} l  l p{0.055 \linewidth}p{0.055 \linewidth}p{0.055 \linewidth}p{0.055 \linewidth}p{0.05 \linewidth}p{0.055 \linewidth}p{0.055 \linewidth}p{0.055 \linewidth}p{0.07 \linewidth} p{0.055 \linewidth} }
\toprule \\
\textbf{Species} $i\,=\,\ldots$ & \multicolumn{1}{c}{\textbf{Calculation method}} & \multicolumn{8}{c}{\textbf{Coexistence mechanisms}} & \multicolumn{2}{c}{\textbf{Invasion growth rate}}\\
& & \multicolumn{4}{c}{small-noise} & \multicolumn{4}{c}{exact} & \multicolumn{1}{c}{approx} & \multicolumn{1}{c}{exact} \\ 
\cmidrule(lr){3-6} \cmidrule(lr){7-10} \cmidrule(lr){11-11} \cmidrule(lr){12-12} \\
 & & $r_i^\prime$or $\Delta E_i$ & $\Delta \rho_i$ & $\Delta N_i$ & $\Delta I_i$ & $r_{i}^{\prime (e)}$or $\Delta E_i^{(e)}$ & $\Delta \rho_{i}^{(e)}$ & $\Delta N_{i}^{(e)}$ & $\Delta I_{i}^{(e)}$ & $\approx \overline{r_i^{\{-i\}}}$ & $\overline{r_i^{\{-i\}}}$ \\ 
\midrule \\
 1 & \text{Scaling factors} & 0.256 & 0.000 & -0.005 & 0.001 & 0.369 & 0 & -0.002 & -0.000
   & 0.252 & 0.367 \\
 1 & \text{Simple comparison} & 0.020 & 0.242 & -0.022 & 0.004 & 0.019 & 0.354 & -0.012 &
   0.005 & 0.244 & 0.367 \\
 1 & \text{Speed conversion factors (generation)} & -0.025 & 0.320 & -0.113 & 0.021 &
   -0.027 & 0.420 & -0.061 & 0.035 & 0.203 & 0.367 \\
 1 & \text{Speed conversion factors (universal)} & -0.039 & 0.344 & -0.140 & 0.026 &
   -0.042 & 0.440 & -0.076 & 0.045 & 0.191 & 0.367 \\
 1 & \text{Invader--invader comparison} & 0 & 0.267 & -0.017 & 0.000 & 0.001 & 0.380 &
   -0.012 & -0.002 & 0.250 & 0.367 \\
   \midrule \\
 \text{2 $\&$ 3} & \text{Scaling factors} & -0.001 & 0.000 & 0.000 & 0.008 & -0.001 & 0 &
   -0.000 & 0.014 & 0.007 & 0.013 \\
 \text{2 $\&$ 3} & \text{Simple comparison} & -0.010 & 0.016 & 0.002 & 0.004 & -0.010 &
   0.015 & 0.001 & 0.007 & 0.012 & 0.013 \\
 \text{2 $\&$ 3} & \text{Speed conversion factors (generation)} & 0.002 & -0.003 & 0.010 &
   0.004 & 0.003 & -0.002 & 0.006 & 0.007 & 0.013 & 0.013 \\
 \text{2 $\&$ 3} & \text{Speed conversion factors (universal)} & 0.003 & -0.004 & 0.010 &
   0.004 & 0.003 & -0.003 & 0.006 & 0.007 & 0.013 & 0.013 \\
 \text{2 $\&$ 3} & \text{Invader--invader comparison} & 0 & 0.005 & 0.001 & 0.005 & -0.001
   & 0.003 & 0.002 & 0.008 & 0.010 & 0.013 \\
\bottomrule \\ 
\end{tabular}
\hspace*{-2cm}
\end{table}

\newpage

\section{Discussion}
\label{Conclusions}

If we can define coexistence mechanisms as measures of the importance of various explanations for coexistence, then they can be straightforwardly used to infer how species are coexisting in real communities (through the analysis of empirically-calibrated models).  In this paper, we have discussed four definitions of coexistence mechanisms, each respectively based on \textit{scaling factors, a simple comparison, speed conversion factors, and an invader--invader comparison}. 

Scaling factors can be useful in theoretical research, but they are not recommended for the purpose of quantifying coexistence mechanisms in real communities. There are better alternative methods for computing coexistence mechanisms (namely the simple comparison and speed conversion factors), each with strengths and weaknesses (Table \ref{procon}). The simple comparison method is easy to compute and interpret, but may give unintuitive results when species have dissimilar generation times. The speed conversion factors work well when species have dissimilar generation times, but are not always well-defined. The invader--invader comparison directly measures the causal effects of low density, but it does not always exist; when it does exist, it does not always quantify the notion of specialization/differentiation. Though we have substantial conceptual arguments in favor of the speed conversion factors, we have only compared methods in two case studies (Section \ref{Case studies}). It is hard to advise the universal adoption of one method. 

We do, however, give the following tentative recommendation: Calculate coexistence mechanism using both the simple comparison method and the scaling factor method. If the results of the two methods are qualitatively similar, great. If the results are qualitatively different, and if species have dissimilar population-dynamical speeds, then place more credence in the speed conversion factors. If species have similar population-dynamical speeds, but the simple comparison method and the speed conversion factor method give qualitatively different inferences, then this uncertainty should be reported. The invader--invader comparison is not recommended, primarily because the method cannot be applied when invasion growth rates are negative.

In the end, the decision between the simple comparison and speed conversion factors may be inconsequential: coexistence is most often studied in guilds of species that have similar generation times, because the putative coexistence of species with similar life histories is more surprising in light of the competitive exclusion principle (\cite{GauzeG.F.GeorgiĭFrant︠s︡evich1934TsfeST}; \cite{levin1970community}). Even when there are small between-species differences in generation times, any error that results from selecting the simple factors over the speed conversion factors will likely be small, relative to the error which results from model misspecification, or failing to account for parameter uncertainty (if one does not calculate coexistence mechanisms across either the joint posterior or bootstrap distribution of model parameters) and structural uncertainty (if one does not calculate coexistence mechanisms for several disparate models). It is important to keep in mind that there are many ways in which a MCT analysis can be provisional.

The simple comparison method captures the notion of specialization, but also captures intrinsic between-species differences that have little to do with coexistence, such as population-dynamical speed. The invader--invader comparison, on the other hand, isolates the effects of rarity, but does not necessarily capture the notion of specialization. We may think of the speed conversion factors as giving the best of both worlds: reducing between-species differences that are irrelevant to coexistence, but capturing the notion of specialization (by remaining within the paradigm of invader--resident comparisons). In fact, the speed conversion factors can be thought of as partially correcting for between-species differences in average fitness (Appendix \ref{speed conversion factors correct for "average fitness differences"}).

We have criticized the scaling factors, the simple comparison, and the invader--invader comparison on the grounds that they can lead to counterintuitive conclusions about how species are coexisting. Our intuitions are rooted in a belief that coexistence mechanisms should measure explanations for coexistence, and that in turn, explanations for coexistence should involve specialization/differentiation that lead to systematic (i.e., across species) rare-species advantages. Historically, the notion of specialization has been central to explanations for coexistence. For example, the heuristic "\ldots each species must consume proportionately more of the resource that more limits its growth" (\cite[p.~96]{tilman1982resourceST}) contains the word "proportionately", which insinuates a cross-species comparison: demographic parameters from multiple species must be considered simultaneously. Indeed, this can be seen in either the mathematical (\cite[p.~77]{tilman1982resourceST}), or the graphical (\cite{chase2003ecological}) versions of Tilman's coexistence theory.

Unfortunately, heady concepts like \textit{specialization} or \textit{population-dynamical speed} do not have formal definitions that apply generally (i.e., in an arbitrary model), meaning that speed conversion factors cannot be justified with a single concise argument. Instead, we have evaluated methods through conceptual analysis and the probing of particular models. Though the two case studies support the usage of speed conversion factors, it is the possible that they can be problematic in contexts that we have failed to imagine. MCT is powerful because it is a general framework: take an arbitrary model, select a few equilibrium parameters, and algorithmically partition the invasion growth rate. In a sense, calculating coexistence mechanisms is the easy part of MCT. The hard part is determining the meaning of the numbers that MCT spits out.

\begin{longtable}[h]{ p{0.2 \linewidth} p{0.4\linewidth} p{0.4\linewidth}}
\caption{Pros and cons of methods for calculating coexistence mechanisms} \\
\label{procon}
Method & Pros & Cons \\
\toprule
\underline{Scaling factors} & & \\ 
 & \tabitem Eliminates $\Delta \rho_i$ when there are more residents than regulating factors, showing that not all species can coexist via classical mechanisms & \tabitem Eliminates $\Delta \rho_i$ when there are more residents than regulating factors; prevents us from determining the degree to which species coexist via classical mechanisms \\
 & \tabitem Converts the units of resident growth to that of invader growth; Useful if species are measured in different units & \tabitem Modulates other coexistence mechanisms, sometimes leading to counterintuitive inferences about how species are coexisting \\
 & & \tabitem Are not uniquely determined when there are more regulating factors than resident species \\
 & & \tabitem Are not uniquely determined (even if there are less regulating factors than resident species) when the invader's $\mathscr{C_j}$ cannot be written as a function of the resident $\mathscr{C_s}$'s (see \eqref{a6}, Section \ref{Scaling factors}) \\
 & & \tabitem Can be sensitive to small changes in inputs if species responses to regulating factors are nearly linearly dependent \\
 & & \tabitem In models with mutualism, can turn an invader--resident difference into an invader--resident sum; the interpretation of coexistence mechanisms as a rare-species advantage is lost \\ 
 \underline{Simple comparison} & & \\ 
 & \tabitem Solves all of the scaling factor cons & \tabitem If some species have fast population dynamics, they will dominate the invader--resident comparison, leading to counterintuitive inferences about how species are coexisting \\
 & \tabitem Easy to compute; works universally & \\
 \underline{Speed conversion factors} & & \\ 
 & \tabitem Solves all of the scaling factor and simple comparison cons; puts species with different generation times on an equal footing & \tabitem If it is not clear what demographic parameters are associated with birth vs. death processes, then generation time is ambiguous \\
 & & \tabitem  Justified by the idea that population-dynamical speed is often irrelevant for coexistence, but this is not always true  \\
 \underline{Invader--invader comparison} & & \\
 & \tabitem Has a straightforward causal interpretation as the effects of perturbing a species to low density (mediated through mean regulating factors, variation in regulating factors, etc.) & \tabitem Can't be computed if the focal species has a negative invasion growth rate or in the case of \textit{the resident strikes back} \\
 & \tabitem No need to worry about scaling factors vs. simple comparison vs. speed conversion factors & \tabitem Can't always be interpreted in terms of specialization \\
\bottomrule
\end{longtable}

\section{Appendixes}

\subsection{Scaling factors in the case of diffuse competition}
\label{Scaling factors in the case of diffuse competition}

Consider the $S$-species Lotka-Volterra model:
\begin{equation}
\frac{1}{n_j(t)} \frac{d n_j(t)}{d t} = k_j - \sum_{k = 1}^S \alpha_{jk} n_{k}(t), \quad j = (1, \ldots, S).
\end{equation}

In the case of diffuse competition where $x$ is interspecific competition and $c$ is intraspecific competition, the competition coefficients are

\begin{equation} 
\alpha_{jk} = 
\begin{cases}
        c & \text{if} \; j = k \\
        x & \text{if} \; j \neq k \\
\end{cases}.
\end{equation}

The regulating factors in this model are the species densities. The matrix of sensitivities to regulating factors (introduced in \eqref{vector_matrix_qir}, Section \ref{Scaling factors}) is thus

\begin{equation}
\boldsymbol{\Phi} = \begin{bmatrix}
c & x & \dots & x \\
x & c & \dots & x \\
\vdots & \vdots & \ddots & \vdots \\
x & x & \dots & c \\
\end{bmatrix}.
\end{equation}

When a species is placed in the invader state, one of the regulating factors vanishes. Therefore, we invert the matrix of responses with the $i$-th row and $i$-th column removed. This matrix, which we will still call $\boldsymbol{\Phi^{\{-i\}}}$, is a square $(S-1) \times (S-1)$ matrix. Note that this is a slight abuse of notation, since in the main text, $\boldsymbol{\Phi^{\{-i\}}}$ represents a matrix where only the invader's \textit{row} has been removed.

To invert $\boldsymbol{\Phi^{\{-i\}}}$, we first decompose it:

\begin{equation}
    \boldsymbol{\Phi^{\{-i\}}} = A + u v^\intercal,
\end{equation}

where $A = \begin{bmatrix}
c-x & 0 & \dots & 0 \\
0 & c-x & \dots & 0 \\
\vdots & \vdots & \ddots & \vdots \\
0 & 0 & \dots & c-x \\
\end{bmatrix}$, $u = \begin{bmatrix}
x \\
\vdots \\
x \\
\end{bmatrix}$, and $v^\intercal = \begin{bmatrix}
1 \dots 1 \\
\end{bmatrix}$. 

The variables $A$, $u$, and $v$ all have $S-1$ rows. Now we can use the Sherman-Morrison formula for the matrix inverse, which states that

\begin{equation}
    \boldsymbol{\Phi^{\{-i\}}} = \left(A + u v^\intercal\right)^{-1} = A^{-1} - \frac{A^{-1}  u v^\intercal A^{-1} }{1 + v^\intercal A^{-1} u}.
\end{equation}

Combined with the well-known fact that the inverse of a diagonal matrix is the matrix of the reciprocals of diagonal elements, we find that
$\left(\boldsymbol{\Phi^{\{-i\}}}\right)^{-1}$ is a symmetric matrix with diagonal elements equal to $\frac{x}{(c-x)(c+x(S-2)}$, and off-diagonal elements equal to $\frac{1}{c-x} - \frac{x}{(c-x)(c+x(S-2)}$.

Since $\boldsymbol{\Phi^{\{-i\}}}$ is invertible, the solution for the scaling factors, \eqref{qir_sol} (Section \ref{Scaling factors}), becomes 

\begin{equation} 
    \boldsymbol{q_{i*}} = \boldsymbol{\Phi}_{i*} \left( \boldsymbol{\Phi}^{(-i)} \right)^{-1}.
\end{equation}

Performing this computation with $\boldsymbol{\Phi}_{i*} = \begin{bmatrix} x \dots x \\ \end{bmatrix}$, a row vector with $S-1$ elements, we get 

\begin{equation}
  q_{is} = \frac{x}{c + (S-2) x}.
\end{equation}

\subsection{Speed conversion factors correct for \textit{average fitness differences}}
\label{speed conversion factors correct for "average fitness differences"}

One side of Modern Coexistence Theory (MCT) is concerned with partitioning the invasion growth rate into coexistence mechanisms like relative nonlinearity, the storage effect, etc. There is an entirely different side of MCT, which is concerted with explaining the coexistence in terms of equalizing mechanisms and stabilizing mechanisms (\cite{chesson1990macarthur}; \cite{chesson2000mechanisms}; \cite{Chesson2018}). \textit{Equalizing mechanisms} weaken competitive differences between species, whereas \textit{stabilizing mechanisms} strengthen niche differences. The important insight here is that different kinds of between-species differences can have different effects on coexistence.

The equalizing vs. stabilizing paradigm only applies to two-species models with Lotka-Volterra-like dynamics (There is a multi-species theory, but the mathematical objects are different; \cite{song2019consequences}). Consider the following parameterization of the Lotka-Volterra Model:

\begin{equation}
    \frac{1}{N_j} \frac{dN_j}{dt} = b_j \left(1- \sum_{k = 1}^2 \alpha_{jk} N_k \right).
\end{equation}

The conditions for coexistence are described by the relation

\begin{equation}
\rho <  \frac{\kappa_1}{\kappa_2} < \frac{1}{\rho},
\end{equation}

where $\rho$ is the \textit{niche overlap} and $\kappa_1 / \kappa_2$ is the \textit{average fitness ratio}. They are defined as 

\begin{equation}
    \rho = \sqrt{\frac{b_1 \alpha_{12} b_2\alpha_{21}}{b_1 \alpha_{11} b_2 \alpha_{22}}}, \quad and
\end{equation}

\begin{equation}
    \frac{\kappa_1}{\kappa_2} = \frac{b_2}{b_1} \sqrt{\frac{\alpha_{21} \alpha_{22}}{\alpha_{11} \alpha_{12}}}.
\end{equation}

The reciprocal of the average fitness ratio, $\kappa_2 / \kappa_1$ includes speed conversion factor, $b_1 / b_2$.  Crucially, the speed parameters, $b_j$, cancel out in the niche overlap, but not the average fitness difference. Therefore, if one accepts that $\kappa_j$ can be rightfully called the average fitness of species $j$ (justification in \cite{Chesson2018}; counterpoint in \cite{barabas2018chesson}), then it is reasonable to think of the action of the speed conversion factors as virtually reducing average fitness differences between species. Even though the average fitness ratio and the speed conversion factors are not identical, dividing by the fitness ratio and multiplying by the speed conversion factors will have largely the same effect if species have different dynamical speeds, but similar competitive effects. This supports our claim (in the main text, Section \ref{Conclusions}) that speed conversion factors are most useful when species have dissimilar population-dynamical speeds.

\section{References}
\printbibliography[heading=none]

\end{document}